\documentclass[graybox]{svmult}
\usepackage{mathptmx}       % selects Times Roman as basic font
\usepackage{helvet}         % selects Helvetica as sans-serif font
\usepackage{courier}        % selects Courier as typewriter font
\usepackage{makeidx}         % allows index generation
\usepackage{graphicx}        % standard LaTeX graphics tool
\usepackage{multicol}        % used for the two-column index
\usepackage[bottom]{footmisc}% places footnotes at page bottom

\newcommand{\be}{\begin{equation}}
\newcommand{\ee}{\end{equation}}
\newcommand{\bdm}{\begin{displaymath}}
\newcommand{\edm}{\end{displaymath}}
\newcommand{\bea}{\begin{eqnarray}}
\newcommand{\eea}{\end{eqnarray}}
\newcommand{\ba}{\begin{align}}
\newcommand{\ea}{\end{align}}

\makeindex             % used for the subject index
\begin{document}

\title{Multidimensional relativistic MHD simulations of Pulsar Wind Nebulae: dynamics and emission}
\titlerunning{Multidimensional relativistic MHD simulations of PWNe: dynamics and emission} 

\author{Luca Del Zanna and Barbara Olmi}
\institute{Dipartimento di Fisica e Astronomia, Universit\`a degli Studi di Firenze, \\
Via G. Sansone 1, 50019 Sesto Fiorentino (FI), Italy. \\ E-mail: \texttt{luca.delzanna@unifi.it}, \texttt{barbara.olmi@unifi.it}} 
%\and Name of Second Author \at Name, Address of Institute \email{name@email.address}}
\maketitle

\abstract{
Pulsar Wind Nebulae, and the Crab nebula in particular, are the best cosmic laboratories to investigate the dynamics of magnetized relativistic outflows and particle acceleration up to PeV energies. Multidimensional MHD modeling by means of numerical simulations has been very successful at reproducing, to the very finest details, the innermost structure of these synchrotron emitting nebulae, as observed in the X-rays. Therefore, the comparison between the simulated source and observations can be used as a powerful diagnostic tool to probe the physical conditions in pulsar winds, like their composition, magnetization, and degree of anisotropy. However, in spite of the wealth of observations and of the accuracy of current MHD models, the precise mechanisms for magnetic field dissipation and for the acceleration of the non-thermal emitting particles are mysteries still puzzling theorists to date. Here we review the methodologies of the computational approach to the modeling of Pulsar Wind Nebulae, discussing the most relevant results and the recent progresses achieved in this fascinating field of high-energy astrophysics.
}
\section{Introduction}
\label{sect:intro}

A pulsar wind nebula (PWN) is a particular class of supernova remnant (SNR) in which the radiation is dominated at all wavelengths by non-thermal mechanisms, namely synchrotron and Inverse Compton (IC) emission by electrons and positrons accelerated to ultra-relativistic velocities. Compared to standard SNRs, their emission is stronger in the central regions rather than at the limbs, this is the reason why PWNe are also known as \emph{plerions}. To date, more than one hundred PWNe have been discovered in our Galaxy and they are among the most luminous sources of the sky in the high-energy bands of emission, X and $\gamma$-rays. Due to its vicinity ($\sim 2$~kpc) and relatively young age ($\sim 1000$ years), the Crab nebula is considered as the PWN prototype, it is certainly the best observed object of its class, and probably the most studied astrophysical source beyond the solar system. Many exhaustive reviews on PWNe have been published in the last decade \cite{Gaensler:2006,Hester:2008,Buhler:2014,Kargaltsev:2015}, thus we refer the reader to these papers for observational details and phenomenology. 

From a theoretical point of view, PWNe represent fantastic astrophysical laboratories, namely for the dynamics of relativistic plasmas, for high-energy conversion mechanisms, and for particle acceleration up to extreme (PeV) energies. They are the best investigated example of \emph{cosmic accelerators}, and almost certainly the principal antimatter factories in the Galaxy. Moreover, PWNe serve as close and well observed benchmarks to test models for similar physics encountered in other classes of sources, like the engines of gamma-ray bursts (GRBs) or active galactic nuclei (AGNs), thus their modeling is of paramount importance for the whole field of high-energy astrophysics. For updated reviews on theoretical modeling of PWNe and open problems subject of current research see \cite{Arons:2012,Amato:2014,Bucciantini:2014e,Olmi:2016}.

As already suggested in the early days of the discovery of pulsars \cite{Pacini:1967,Pacini:1968,Gold:1968}, the Crab nebula and the other PWNe, especially young ones, are very efficiently powered  (up to $30\%$) by the spin-down energy losses from a neutron star (NS) located at their center. PWNe are basically hot bubbles of magnetized plasma, arising from the confinement of the relativistic wind from a rapidly rotating and strongly magnetized pulsar, in the form of leptons and electromagnetic fields, by the external slowly expanding ejecta of the SNR. At the interface between the pulsar wind (PW) and the plerionic nebula, a reverse MHD shock forms, generally referred to as the \emph{termination shock} (TS). This is the location where the wind bulk flow converts into disordered motion (heat), and where pairs (possibly even ions) are accelerated, becoming responsible for the synchrotron and IC non-thermal emission from the nebula. Outside the PWN, the SNR expands in the interstellar medium (ISM) at supersonic but sub-relativistic speed, as a blast wave. 

Time-dependent one-zone models and steady-state relativistic hydro/MHD spherically symmetric models \cite{Pacini:1973,Rees:1974,Kennel:1984,Kennel:1984a} reproduced quite well the observed evolution, spectrum, and global structure of PWNe, thus the theoretical picture arising from these works has been considered as rather successful up to the turn of the century. In Fig.~{\ref{fig:kc} we sketch the classical radially symmetric model of the Crab nebula proposed the cited models, useful to recognize the main components of the PWN/SNR system.

\begin{figure}[t]
%\sidecaption[t]
\centering
\includegraphics[scale=0.4]{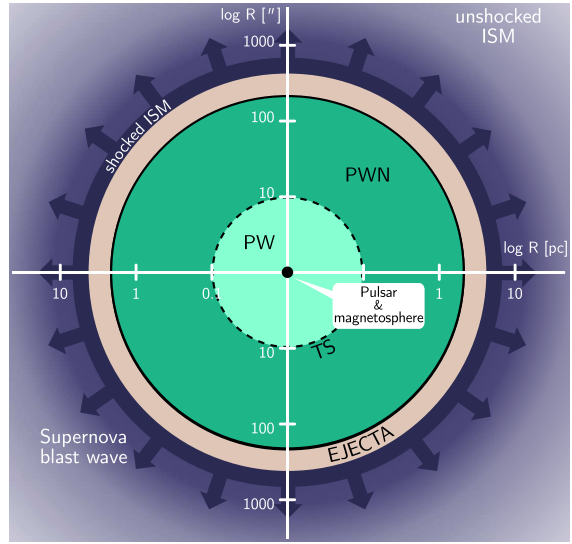} 
\caption{The radial structure of a PWN/SNR system for the case of the Crab nebula, with distances in logarithmic scale expressed in pc and arcsec along the $x$ and $y$ axes, respectively (adapted from \cite{Kennel:1984}). The central region contains the ultra-relativistic, cold pulsar wind (PW), surrounded by the hot bubble of magnetized plasma, that is the PWN itself shining with synchrotron light. The nebula is bounded by the radii of the wind termination shock ($R_\mathrm{TS}\simeq 0.1$~pc) and of the contact discontinuity with the external dense material of the SNR ejecta ($R_\mathrm{PWN}\simeq 2$~pc). Beyond the freely expanding ejecta, the ISM material is shocked by the SNR blast wave propagating outwards.
}
\label{fig:kc} 
\end{figure}

A revival of interest in PWNe came after the first detailed images of the Crab nebula from the X-ray Chandra satellite \cite{Weisskopf:2000}. Its inner structure revealed a wealth of fine details such as bright and variable features, close to the location of the pulsar, and an unexpected axisymmetric \emph{jet-torus} structure. This was later found to be common to all other young PWNe where the inner morphology is discernible, most notably around the Vela pulsar \cite{Helfand:2001,Pavlov:2001,Ng:2004}. The torus of enhanced emission appears to be asymmetric in brightness (a clear signature of the Doppler boosting effect), and the same is true for the two polar jets, mainly aligned with the symmetry axis. The torus is threaded by rings and the innermost one delimits a dark zone of absence of emission. Bright and variable \emph{knots} are observed along the inner ring and close to the pulsar. Moving features with mildly relativistic speeds are observed along the jets (that \emph{kink} at their extremes) and in the equatorial region before the torus, as outgoing arc-like features named \emph{wisps}, also seen in the optical band \cite{Scargle:1969,Hester:2002}.

This situation was felt by theoretical astrophysicists as a new, exciting challenge. In particular, the presence of jets originating very near the pulsar, apparently within the PW itself (identified as the dark region inside the inner ring), appeared puzzling, since plasma collimation by magnetic fields in a relativistic outflow is known to be inefficient. However, provided the torus of enhanced emission could arise from a higher energy input of the wind in the equatorial zone, the TS would then assume an oblate shape, with an inner boundary much closer to the central NS along the axis, depending on the degree of anisotropy in the energy flux \cite{Bogovalov:2002}. Such a situation allows collimation of jets to occur inside the post-shock nebula, where MHD hoop stresses are now efficient due to the mild velocity of the outflow \cite{Lyubarsky:2002}.

The theoretical picture looked promising, though self-consistent calculations were needed to definitively prove it. At that time, only a couple of relativistic MHD codes, needed to simulate the interaction of the PW and the daughter PWN with the external remnants, were available \cite{Komissarov:1999,Del-Zanna:2002,Del-Zanna:2003}. The first axisymmetric simulations \cite{Komissarov:2003,Komissarov:2004a,Del-Zanna:2004}  clearly verified Lyubarsky's idea: given an anisotropic energy flux of the PW, an oblate TS forms and the mildly relativistic post-shock flow is redirected towards the polar axis by magnetic forces acting in the PWN. Moreover, synthetic synchrotron surface brightness maps were able to reproduce not only the major jet-torus inner structure, but also most of the fine details observed in the X-rays \cite{Del-Zanna:2006}. The modeling was later extended to the polarization properties of PWNe \cite{Bucciantini:2005a}, to the $\gamma$-rays produced due to IC scattering  \cite{Volpi:2008c} (by the same ultra-relativistic leptons responsible for synchrotron emission onto various photon background fields), to the detailed modeling of radio emission \cite{Olmi:2014}, to the dynamics of wisps variability in various photon energy bands \cite{Camus:2009,Olmi:2015}, and to the inspection of the role of the pulsar's magnetosphere obliquity on the PWN structure \cite{Buhler:2016}. In spite of its simplicity and of the highly successful results obtained, axisymmetry was then abandoned in favor of full 3-D simulations \cite{Porth:2013,Porth:2014,Olmi:2016}, allowing for a more realistic distribution and dissipation of the nebular magnetic field.

It is not our intention to cover in the present review all topics related to PWN numerical modeling. For instance, PWNe will be used as benchmarks to test future X-ray polarimetry instruments, thus synchtrotron diagnostic tools based on MHD modeling will be certainly needed \cite{Del-Zanna:2006,Bucciantini:2005a}. Inside the hot and magnetized PWN bubble, the development of fluid and MHD instabilities are expected (Kelvin-Helmoltz around the TS, Rayleigh-Taylor at the contact discontinuity with ejecta, kink-like in jets, tearing in thin current sheets), and numerical simulations are a very useful tool to investigate their evolution in the relativistic regime \cite{Bucciantini:2006a,Bucciantini:2004a,Porth:2014a,Mizuno:2011,Mignone:2013,Del-Zanna:2016}. A rich and complex phenomenology comes from the modeling of evolved PWNe interacting with different environments, like the ISM in the case of bow-shock nebulae from pulsars in supersonic motion \cite{Bucciantini:2005} or stellar winds in the case of peculiar binary systems \cite{Bogovalov:2012,Bosch-Ramon:2012,Dubus:2015}. A similar approach together with the tools employed in the investigation of PWNe have been also applied to the modeling of long and short GRBs: in this scenario the central engine is a millisecond proto-magnetar inflating a relativistic magnetized bubble confined by the stellar (or NS-NS merger) external envelope, through which jets collimated by hoop stresses will eventually penetrate, thus sharing a very similar dynamics with PWNe \cite{Komissarov:2007a,Bucciantini:2009,Bucciantini:2012}.

In the following we shall mainly concentrate on the modeling of \emph{standard} (young) PWNe, focussing on multidimensional (both 2-D axisymmetric and full 3-D) relativistic MHD simulations of the Crab nebula and similar objects. First the physical framework and the methods employed in the above cited works will be reviewed, including the recipes for computing the non-thermal emission on top of the numerical results. The main findings and achievements provided by relativistic MHD modeling will be described in separate sections, where (partial) answers to the open issues of PWN physics will be proposed. Finally, conclusions will be drawn and  the possible future of MHD modeling will be discussed.

\section{The setup for numerical modeling}
\label{sect:setup}

As outlined in the introduction and as will be described in details in the remainder of this chapter, the (relativistic) MHD assumption has proved to be very successful for PWNe modeling, starting from the seminal works \cite{Kennel:1984,Kennel:1984a}. We recall that the MHD assumption is valid when: 1) Larmor radii of particles are much smaller than the typical size of the nebula, in this case the TS radius; 2) a one-fluid approximation can be assumed, that is a pure pair plasma is considered, neglecting the possible presence of ions; 3) radiative losses are small, in the sense that only the distribution functions of non-thermal accelerated particles are affected, while preserving a Maxwellian shape for particles composing the bulk plasma (for which the synchrotron lifetime is longer than the age of the nebula). These three conditions appear to be all met in the Crab nebula and similar young sources.

The initialization of any MHD simulation of the PWN/SNR system is basically the same regardless of the symmetries imposed. A relativistic wind with radial flow must be injected from the inner boundary of the numerical box, whereas radially expanding SNR ejecta and a static and unperturbed ISM must be initialized elsewhere. Notice that the PWN itself is not defined at the initial time, while it is expected to be created self-consistently during the computation of the system evolution. In the following we shall discuss in some details the equations, methods, and basic initialization needed to perform relativistic MHD simulations of PWNe.

\subsection{Relativistic MHD and numerical methods}

The full and accurate description of the dynamics of a pulsar wind, from its origin within the NS magnetosphere to its interaction with the supernova remnant (and of the latter with the ISM), is certainly challenging not only for the very different spatial scales (a large factor separates the inner light cylinder radius from the TS radius, namely $R_\mathrm{TS}/R_\mathrm{LC}\sim 10^9$), but also due to the diverse physical regimes involved. This is important from the point of view of numerical modeling, as it is not easy to match different regions where the evolution is described by different set of equations.

The presence of the huge magnetic field anchored to the NS crust, $B\sim 10^{12}~\mathrm{G}$ for standard pulsars and even $B\sim 10^{15}~\mathrm{G}$ for magnetars, and of the electric fields induced by rotation, allows one to employ the \emph{force-free electrodynamics} (FFE) approximation within the magnetosphere: electromagnetic forces dominate and balance themselves, while the plasma inertia and pressure forces may be neglected, as shown in the first 3-D simulations \cite{Spitkovsky:2006}. Moreover, a precise modeling of the NS magnetosphere would require to include general relativistic effects, either in the FFE regime \cite{Petri:2016}, in the MHD fluid approximation \cite{Pili:2014,Bucciantini:2015}, or employing full particle-in-cell (PIC) simulations \cite{Philippov:2015}. In the wind zone, the escaping particles (pairs and maybe protons) are accelerated to relativistic velocities, though their thermal speed remains negligible, so that a cold relativistic fluid-like description is more appropriate \cite{Bogovalov:1999a}. After the TS the plasma is heated to relativistically hot temperatures and the magnetic field is amplified up to equipartition, allowing for a full relativistic MHD description. At the external boundary of the PWN, after the contact discontinuity radius $R_\mathrm{PWN}$, the slowly expanding supernova ejecta follow standard Newtonian hydrodynamics, up to their outer regions where they merge into the static ISM (through a bow-shock transition in the case of pulsars in supersonic motion \cite{Bucciantini:2005}).

The complexity of the physics described above clearly explains why numerical simulations are limited to either just the magnetosphere and the PW launching region, or just to the PWN and to its interaction with the environment. In the latter case, the wind before the TS is prescribed as an inner boundary condition, and the evolution of the PWN system is followed by solving the relativistic MHD equations (in a flat spacetime). Even if the external medium is non-relativistic, the standard equation of state for an ideal gas with constant adiabatic index is typically adopted. 

The equations of (special) relativistic MHD, written in conservative form as appropriate for the use in shock-capturing numerical codes, express the conservation of mass, momentum, and energy as follows:
\be
\frac{\partial}{\partial t} (\rho\gamma) + \nabla \cdot (\rho\gamma\,\vec{v}) =0,
\ee
\be
\frac{\partial}{\partial t}  \left( \frac{w}{c^2}\gamma^2\vec{v} + \frac{\vec{E}\times\vec{B} }{4\pi c}\right) + \nabla \cdot \left[
\frac{w}{c^2}\gamma^2\vec{v}\vec{v} - \frac{\vec{E}\vec{E} + \vec{B}\vec{B} }{4\pi} + \left(P + \frac{E^2+B^2}{8\pi}\right) \vec{I} \right] = 0, 
\ee
\be
\frac{\partial}{\partial t}  \left( w\gamma^2 - P + \frac{E^2+B^2}{8\pi} \right) +  \nabla \cdot  \left[ w\gamma^2\vec{v} + \frac{c}{4\pi}(\vec{E}\times\vec{B}) \right] = 0.
\ee
Here $\rho$ is the rest mass density of the fluid, $P$ the pressure in the local rest frame, $\vec{v}$ is the bulk flow velocity (and $\gamma = 1/\sqrt{1-v^2/c^2}$ is its Lorentz factor), $w=\rho c^2 + 4P$ is the relativistic enthalpy (for a relativistic ideal gas with adiabatic index $4/3$), $\vec{E}$, $\vec{B}$ are the electric and magnetic field as measured in the laboratory frame, $\vec{I}$ is the identity tensor. We recall that in the absence of dissipation, the ideal MHD approximation requires the vanishing of the comoving electric field, that is $\vec{E} = - (1/c)\vec{v}\times\vec{B}$ is used in the above equations. Thus, the electric field is a derived quantity and Maxwell equations just provide the evolution of the magnetic field. In particular, Faraday's law turns into the induction equation
\be
\frac{\partial  \vec{B}}{\partial t} = \nabla \times (\vec{v}\times\vec{B}),
\ee
to be supplemented with the no-monopole constraint $\nabla\cdot\vec{B}=0$.

Modern shock-capturing codes employed for PWN simulations \cite{Komissarov:1999,Del-Zanna:2002,Del-Zanna:2003,Del-Zanna:2007,Mignone:2007a,Mignone:2012a,Porth:2014b} solve the above ideal relativistic MHD equations by using reconstruction-evolution methods based on finite-differences or finite-volumes schemes. Due to the presence of strong shocks and discontinuities, typically the evolution is only second-order accurate in space and time, based on the use of upwind limited reconstruction and Runge-Kutta methods, respectively. The Riemann problem at cell interfaces is usually approximated via a simple and robust two-wave method (see the cited papers for further numerical details), especially in the MHD case. When radially or axially symmetric simulations are performed, spherical coordinates are employed and a (static) grid stretching along the radial direction is enforced to better resolve the delicate wind region and its transition to the inner PWN. In the 3-D case, it is instead convenient to adopt Cartesian coordinates and \emph{adaptive mesh refinement} (AMR) methods (with several levels needed to enforce the ultra-relativistic PW).

\subsection{Model setup and 1-D dynamics}

Any numerical simulation of the PWN system evolution is initialized by providing the wind, which must be also injected during the evolution at a given radius (which acts as the inner boundary condition). In addition, initial conditions beyond that radius must comprise the SNR density and velocity distribution, and eventually those pertaining the ISM in which the ejecta are expanding.

The steady-state structure of a PWN evolving inside a slowly expanding SNR can be modeled analytically by using relativistic MHD. The shock jump conditions and the radial profile of all quantities are seen to depend crucially on the wind magnetization parameter, calculated just ahead of the TS and defined as 
\be
\sigma_0 = \frac{B^2}{4\pi \rho c^2 \gamma^2}.
\label{eq:sigma}
\ee
This is the ratio between Poynting to kinetic energy fluxes in the (cold) relativistic PW, with a Lorentz factor $\gamma\gg 1$ ($v\to c$), density $\rho$, and magnetic field $B$, supposed to be transverse to the flow (only the toroidal component survives at large distances from the pulsar). When 1-D radial models are adopted, the tension of the toroidal field $B$ amplified post-shock builds up for steady energy injection by the PW, and the confinement by the external SNR, supposed to move at velocity $V_\mathrm{PWN}$, requires that the magnetization is small ($\sigma_0\ll 1$). In that limit, assuming $B\sim r^{-1}$ and neglecting energy losses in the PWN, the characteristic radii bounding the PWN and its expansion velocity satisfy the two relations \cite{Rees:1974}
\be
\frac{R_\mathrm{TS}}{R_\mathrm{PWN}} \simeq \sqrt{\frac{V_\mathrm{PWN}}{c}} \simeq \sqrt{\sigma_0}.
\label{eq:ratio}
\ee
In the case of the Crab nebula an expansion velocity of $V_\mathrm{PWN}\simeq 1500$~km/s requires a low magnetization value of $\sigma_0=5\times 10^{-3}$, yielding $R_\mathrm{TS}/R_\mathrm{PWN}\simeq  0.07$, slightly larger compared to what inferred from observations. The refined models in which the magnetic field is retrieved self-consistently from the MHD equations \cite{Kennel:1984} indicate as best-fit values $V_\mathrm{PWN}\simeq 2000$~km/s, $\sigma_0=3\times 10^{-3}$, $R_\mathrm{TS}/R_\mathrm{PWN}\simeq  0.05$ (in accordance with the observed values, see Fig.~\ref{fig:kc}), assuming a constant pulsar spindown luminosity $L_0=5\times 10^{38}$~erg/s.

In the early stage of the evolution, the environment where the PWN bubble forms and propagates mainly consists of stellar ejecta (considered as cold and unmagnetized for simplicity) in free expansion into the ISM. Only later a reverse shock will form and the evolution enters the Sedov-Taylor phase. Spherically symmetric hydrodynamics and relativistic MHD simulations of the PWN/SNR system in the two phases can be found in \cite{van-der-Swaluw:2001,Bucciantini:2003,Bucciantini:2004}. Most of the multidimensional MHD simulations of PWNe are usually limited to the initial free expansion phase (as relevant for the Crab nebula), and assume the same self-similar model for the cold ejecta as adopted in the cited works. This solution entails a Hubble-type profile for the velocity, a spatially constant density, and for $r\le r_\mathrm{ej}(t_0)=v_\mathrm{ej}\, t_0$ ($r_\mathrm{ej}$ is the external radius of the SNR at time $t=t_0$ after the explosion) we have
\be
v(r,t_0) = v_\mathrm{ej} \frac{r}{r_\mathrm{ej}(t_0)}, \quad
\rho_\mathrm{ej} (t_0) = \frac{3}{4\pi} \frac{M_\mathrm{ej} }{r_\mathrm{ej}^3(t_0)},  \quad
v_\mathrm{ej} = \sqrt{ \frac{10}{3}\frac{E_\mathrm{ej}}{M_\mathrm{ej}} },
\label{eq:ejecta}
\ee
where the two latter quantities have been determined by computing the total mass and kinetic energy of the ejecta, respectively. Typical values commonly assumed for 1-D as well as multidimensional numerical models of the Crab nebula are $M_\mathrm{ej} = 3M_\odot = 6\times 10^{33}$~g and $E_\mathrm{ej} = 10^{51}$~erg \cite{Del-Zanna:2004,Porth:2014}. Once $t_0$, or $r_\mathrm{ej}(t_0)$, is chosen, the above settings define the ambient initial condition in which the nascent PWN will propagate at later times of evolution $t>t_0$. For $r>r_\mathrm{ej}(t_0)$ standard static ISM conditions can be applied or, if the outer blast wave is not of primary interest, the ejecta solution can be extrapolated up to the external computational boundaries. 

\begin{figure}[t]
%\sidecaption[t]
\centering
\includegraphics[scale=0.4]{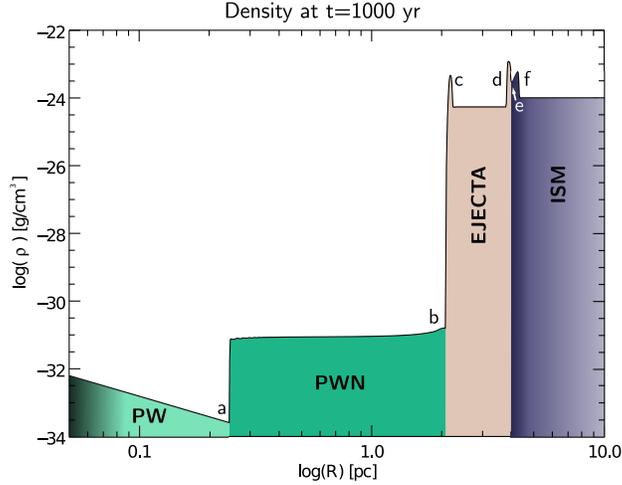} 
\caption{The mass density profile from radially symmetric relativistic MHD simulations of the Crab nebula at the present age of $t=1000$~yr. The adopted parameters are $L_0=5\times 10^{38}$~erg/s, $\sigma_0=3\times 10^{-3}$, $M_\mathrm{ej} = 3M_\odot = 6\times 10^{33}$~g and $E_\mathrm{ej} = 10^{51}$~erg. The PW is shocked at $a$ where the hot PWN is created, $b$ is the contact discontinuity separating the leptonic wind plasma from the swept-up shell of ejecta material. The constant density profile of the SNR ejecta is apparent from $c$ and $d$, where the latter is the position of the reverse shock arising from interaction with ISM. Beyond the contact discontinuity in $e$ we have the shocked ISM, propagating through the blast wave in $f$ into the unperturbed medium with standard conditions $\rho=10^{-24}$~g~cm$^{-3}$ and $T=10^4$~K.}
\label{fig:1D} 
\end{figure}

The radial structure of the PWN/SNR system as arising from 1-D relativistic MHD simulation is shown in Fig.~\ref{fig:1D}, where we adopt the same notation and colors as in the sketch of Fig.~\ref{fig:kc} and the same labeling for discontinuities as in \cite{Bucciantini:2003}. The mass density is plotted in logarithmic scale, for the already cited settings relevant for the Crab nebula as employed in semi-analytical models \cite{Kennel:1984}, and for the ejecta profile in Eq.~(\ref{eq:ejecta}). The initial decrease as $\sim r^{-2}$ is the PW zone, the plateau between $a$ (the TS) and $b$ (the contact discontinuity) is the hot PWN. Notice the huge density jump between the light leptonic material and the swept-up shell of dense ejecta, confined by the shock in $c$, which is challenging to handle even for shock-capturing schemes. Given the self-similar solution for the ejecta employed here (and for constant $L_0$), the PWN is expected to expand as $R_\mathrm{PWN}(t)\sim t^{6/5}$ \cite{Reynolds:1984}. The external SNR blast wave propagates into the unperturbed ISM through the shock in $f$, heating the material and driving a shock backwards from $e$ to $d$ into the constant-density ejecta. When this reverse shock reaches the PWN, the free expansion stage ends, and after a transition phase in which ejecta are fully heated the SNR expansion enters the Sedov-Taylor stage, appropriate for the evolution of old PWN/SNR systems (e.g. the Vela nebula), not treated here.

\subsection{Treatment of non-thermal emission}
\label{sect:sync}

Here we discuss how to extract synthetic synchrotron surface brightness maps and integrated spectra based on the results of multidimensional MHD simulations. While sophisticated methods relying on the kinetic transport of the emitting particles could be implemented \cite{Bai:2015,Porth:2016}, simpler recipes that still take into account particles advection, adiabatic losses and synchrotron radiative losses are more widespread in the community. These simplified (though less accurate) methods rely on the use of \emph{tracers} to be added as new evolution equations to the MHD system. Notice that the back reaction of energy losses of particles on the fluid is neglected within this simplified approach. A full description of the method can be found in \cite{Del-Zanna:2006}, later extended to gamma-ray IC emission \cite{Volpi:2008c}, and further refined \cite{Camus:2009,Olmi:2014}.

The broadband, non-thermal spectrum of PWNe, and of the Crab nebula in particular, is known to be best reproduced if the contribution by two families of emitting electrons are considered: one responsible for the low energy emission (radio particles) and one for the high energy emission (optical/X-ray particles) \cite{Atoyan:1996}. While radio electrons have a very long radiative life and can be, indifferently, either evolved or uniformly distributed throughout the nebula \cite{Olmi:2014}, the X-ray family must be injected at the TS and evolved accurately, since synchrotron burn-off is expected to result in the diverse appearance of PWNe at different photon energies. In addition, different species of electrons may be accelerated at different locations along the TS, so that it could be necessary to evolve more than one family of emitting particles \cite{Olmi:2015}.

The form of the distribution function of freshly injected particles is usually taken as a power-law of index $p$ (an exponential cut-off may be added), that is
\be
f_0 (\epsilon_0) \propto \epsilon_0^{-p}, \quad \epsilon_0^\mathrm{min} \leq \epsilon_0 \leq \epsilon_0^\mathrm{max},
\label{eq:f0}
\ee
where $\epsilon_0$ is the particle energy, here in units of $mc^2$ (that is its Lorentz factor). The local distribution function of wind particles, at any location and time of the PWN evolution, is determined by the conservation of particle number along the streamlines, taking into account adiabatic and synchrotron losses as anticipated. If we indicate with $n$ the number density, for each species of accelerated particles, and with $n_0$ its value immediately downstream of the shock, the result is
\be
f(\epsilon) = \left( \frac{n}{n_0} \right)^{4/3} \left( \frac{\epsilon_0}{\epsilon} \right)^2 f_0 (\epsilon_0), \quad
\epsilon_0 = \left( \frac{n}{n_0} \right)^{-1/3} \frac{\epsilon}{1-\epsilon/\epsilon_\infty},
\label{eq:f}
\ee
where $n_0 = \int_{\epsilon_0^\mathrm{min}}^{\epsilon_0^\mathrm{max}}  f_0 (\epsilon_0) d\epsilon_0$.

In the above expressions, the factors with $n/n_0$ are those taking into account the \emph{adiabatic} changes in volume of the fluid element. These quantities are evolved together with the MHD equations, in conservative form, as
\be
\frac{\partial}{\partial t} (n\gamma) + \nabla\cdot (n \gamma\, \vec{v}) = 0, \quad
\frac{\partial}{\partial t} (n_0 \,n\gamma) + \nabla\cdot (n_0 \,n \gamma\, \vec{v}) = 0, 
\ee
so that $n$ obeys to the same continuity equation as $\rho$, while $n_0$ is simply advected along streamlines after its definition post-shock of the TS. A simplified version of this procedure \cite{Del-Zanna:2006}, only feasible for $p\simeq 2$, is to neglect the volume changes in the above expression for $\epsilon_0$ and to replace the $(n/n_0)^{4/3}$ term with a term proportional to the fluid pressure $P$, though in this way we lose the information about where a particle has been injected. Notice that if different families of particles are injected at the TS, two more evolution equations must be solved for each additional species.

A last evolution equation must be introduced for the quantity $\epsilon_\infty$, which represents the highest attainable (normalized) energy that any particle may have at any instant and location within the PWN, to be initialized to a large number at injection (say $> 10^{10}$). The ratio $\epsilon/\epsilon_\infty  < 1$ provides the radiative losses by modifying the distribution function as in Eq.~(\ref{eq:f}). The equation for $\epsilon_\infty$ is
\be
\frac{\partial}{\partial t}( \rho^{2/3} \epsilon_\infty \gamma)+\nabla\cdot 
(\rho^{2/3} \epsilon_\infty \gamma\,\vec{v})
= - \frac{4e^4 {B^\prime}^2}{9m_\mathrm{e}^3c^5}  \rho^{2/3} \epsilon^2_\infty,
\label{eq:infty}
\ee
where the source term is due to synchrotron losses, $B^\prime$ is the magnetic field strength in the fluid rest frame, and isotropy in pitch angle distribution has been assumed. 

The local emissivity function, at any time and position in the nebula, is given by the integral (primed quantities refer to the fluid rest frame)
\be
j^\prime_\nu(\nu^\prime,\vec{n}^\prime)= \!\!\int 
\frac{2e^4}{3m_\mathrm{e}^2c^3} |\vec{B}^\prime\times\vec{n}^\prime|^2\epsilon^2 
\mathcal{S}(\nu^\prime,\nu^\prime_c) 
f(\epsilon)\mathrm{d}\epsilon, 
\label{eq:j1}
\ee
where $\vec{n}$ is the direction pointing towards the observer (normal to the plane of the sky) and where the spectral density function $\mathcal{S}$ for synchrotron radiation can be either computed exactly \cite{Rybicki:1979}, or it can be approximated to a delta function of the critical emission frequency
\be
\nu^\prime_c(\epsilon) = \frac{3e}{4\pi m_\mathrm{e}c}\,|\vec{B}^\prime\times\vec{n}^\prime|\epsilon^2,
\ee
so that the integral above can be calculated analytically. In order to obtain the emissivity in the observer's fixed frame of reference, relativistic corrections must be taken into account, by computing $j_{\nu}=D^2 j_{\nu}^\prime$ and $\nu=D\nu^\prime$ in Eq.~(\ref{eq:j1}), where $D=[\gamma \,(1-\vec{v}\cdot\vec{n}/c)]^{-1}$ is the Doppler boosting factor. 

In the particular case where the power-law form in Eq.~(\ref{eq:f0}) is used for particles injected at the TS, and if monochromatic emission is assumed for simplicity, the spectral emissivity function in the laboratory frame becomes
\be
j_\nu (\nu,\vec{n}) \propto n_0 (n/n_0)^{\frac{2+p}{3}} D^\frac{3+p}{2}\,|\vec{B}^\prime\times\vec{n}^\prime|^\frac{1+p}{2}( 1 - \sqrt{\nu/\nu_\infty})^{p-2}\, \nu^{-\frac{p-1}{2}}.
\ee
Here $\nu_\infty = D \nu^\prime_c (\epsilon_\infty)$ is the cut-off frequency for synchrotron burn-off, and the emissivity will be different from zero only where $\nu < \nu_\infty $. Additional details and the recipes to compute surface brightness maps, spectral index maps, polarization maps, and integrated spectra from the emissivity $j_\nu$ can be found in \cite{Del-Zanna:2006}.

\section{The flow dynamics and the jet-torus inner structure}

Let us now discuss the main results found by the numerical community in more than a decade of relativistic MHD simulations of PWNe. The first issue is certainly the inner jet-torus structure revealed by X-ray observations.

The detailed angular dependence of the PW is crucial to determine the shape that the PWN will assume during the evolution in multidimensional simulations. The energy flux is usually initialized as axially symmetric (with respect to the pulsar rotation axis), even in 3-D simulations, and it must be taken anisotropic in order to reproduce the observed jet-torus inner structure \cite{Lyubarsky:2002}. This behavior is roughly that predicted by the split-monopole model of aligned rotators \cite{Michel:1973}, where a radiation-like toroidal component $B \equiv B_\phi\sim r^{-1}\sin\theta$ dominates at large distances from the light cylinder, for updated reviews on pulsar magnetospheres and their winds see \cite{Petri:2016a,Cerutti:2016}. The angular dependence of a radial, ultra-relativistic, cold wind ($v\equiv v_r \simeq c, \, \gamma=\sqrt{1-v^2/c^2}\gg 1, \, P\ll \rho c^2$) of constant spindown luminosity $L_0$ can be thus introduced as
\be
L_0 \mathcal{F}(\theta) = 4\pi r^2 c \left( \rho c^2 \gamma^2 + \frac{B^2}{4\pi} \right) =  
4\pi r^2 \rho c^3 \gamma^2  [1+\sigma(\theta)],
\ee
where $\textstyle{\frac{1}{4\pi}} \int \mathcal{F}(\theta) d\Omega = 1$ and $\sigma(\theta)$ is the magnetization function. Once $\mathcal{F}(\theta)$ and $\sigma(\theta)$ are chosen, the angular dependence of $\rho\gamma^2$ and $B^2$ (both decaying as $r^{-2}$) are also known and the wind is fully characterized (equivalently, one can provide the angular shape of $B$ directly). The wind Lorentz factor can be either taken as constant or dependent on $\theta$, assuming an isotropic mass flux in this latter case \cite{Del-Zanna:2004}, in any case limited to the range $\gamma = 10-100$ to avoid computational problems. 

Various choices are possible for the two angular functions. The radially symmetric case is retrieved with $\mathcal{F}=1$ and $\sigma=\sigma_0$, where the latter is the usual pulsar wind magnetization parameter of Eq.~(\ref{eq:sigma}) defined in \cite{Kennel:1984}. Already in the very first axisymmetric simulation \cite{Komissarov:2003} the toroidal magnetic field was shaped to allow for a striped wind region of lower magnetization around the equatorial plane, see also \cite{Del-Zanna:2004,Olmi:2014} for the alternative definitions used by our group. Typical settings, also employed in recent 3-D simulations \cite{Olmi:2016}, define a wind energy flux which is 10 times higher in the equatorial direction with respect to the polar ones, and a magnetization function shaped as in Fig.~\ref{fig:sigma}, depending on $\sigma_0$ ($B\propto\sqrt{\sigma_0}$) and another parameter $b$ providing the size of the equatorial striped region, in turn depending on the pulsar's magnetosphere inclination angle  ($b\to\infty$ for the split-monopole solution without the equatorial region of lower magnetization). More complex assumptions, also valid for strongly inclined rotators, can be found in \cite{Komissarov:2013,Porth:2014,Philippov:2015,Tchekhovskoy:2016}.

\begin{figure}[t]
\centering
\includegraphics[scale=0.4]{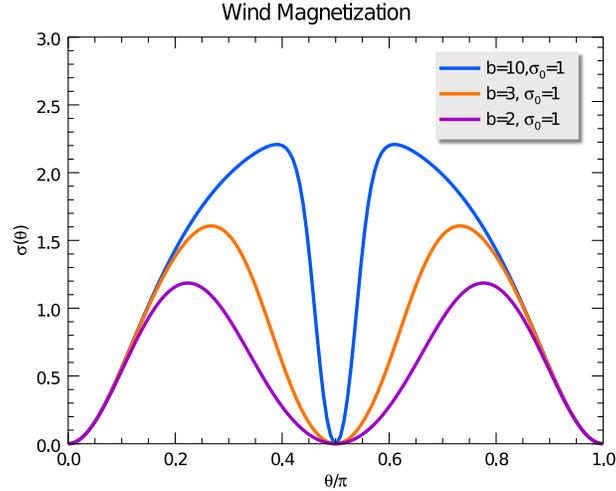} 
\caption{The angular dependence of the wind magnetization function $\sigma(\theta)$, here for the reference value of $\sigma_0=1$ employed in \cite{Olmi:2016}, from which the figure is taken, and three values of the parameter $b$, shaping the toroidal magnetic field \cite{Del-Zanna:2004,Olmi:2014}. The upper (blue) curve refers to the narrow striped-wind region case, used by the authors to reproduced the Crab nebula observations, the bottom (purple) case roughly reproduces the choice in the B3D run by \cite{Porth:2014}.}
\label{fig:sigma} 
\end{figure}

The typical flow structure arising in axisymmetric simulations in the cases of low and intermediate magnetization is shown in Fig.~\ref{fig:flow}, taken from \cite{Del-Zanna:2004} to which the reader is referenced for further details. In the left panel we show the flow and shock pattern arising around the TS in the upper meridional plane (here for a run without the striped wind region and imposing equatorial symmetry). Labels refer to: A) ultra-relativistic wind region; B) subsonic equatorial outflow; C) equatorial supersonic funnel; D) super-fastmagnetosonic shocked outflow; a) termination shock (TS) front; b) \emph{rim} shock; c) fastmagnetosonic surface. The magnetization parameter is that assumed in spherical and stationary models $\sigma_0=0.003$, though, due to the anisotropic structure of the injected PW, only in the equatorial region we observe the expected transition to a shocked flow speed of $v\simeq c/3$ (region B). A complex shock pattern arises away from the equatorial belt, and a fast-speed channel with $v\simeq 0.7$ forms along the TS boundary from the poles (region D) and survives at large distances bending towards the equator (region C).

\begin{figure}[t]
%\sidecaption[t]
\centering
\includegraphics[height=57mm]{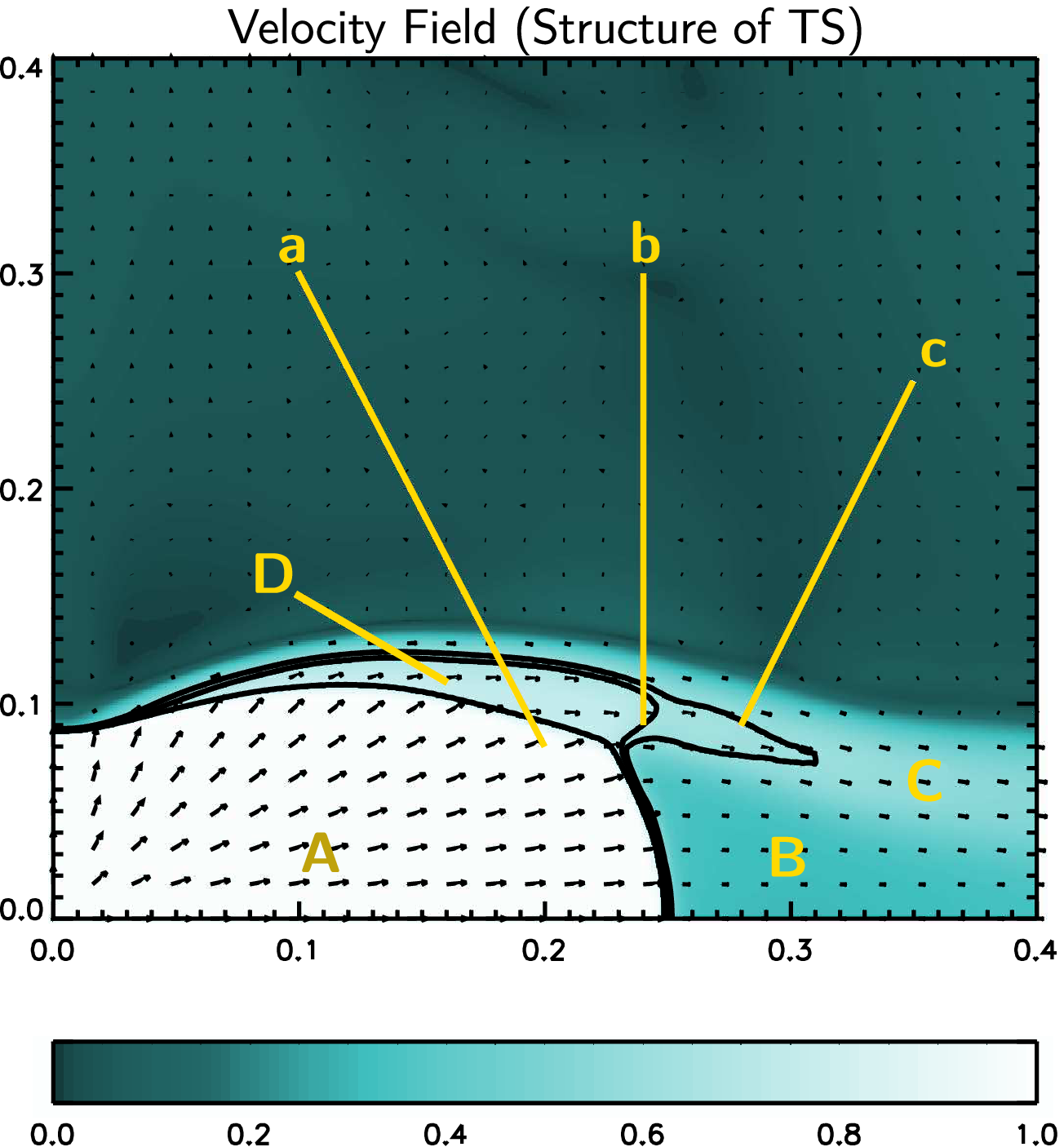} \qquad
\includegraphics[height=57mm]{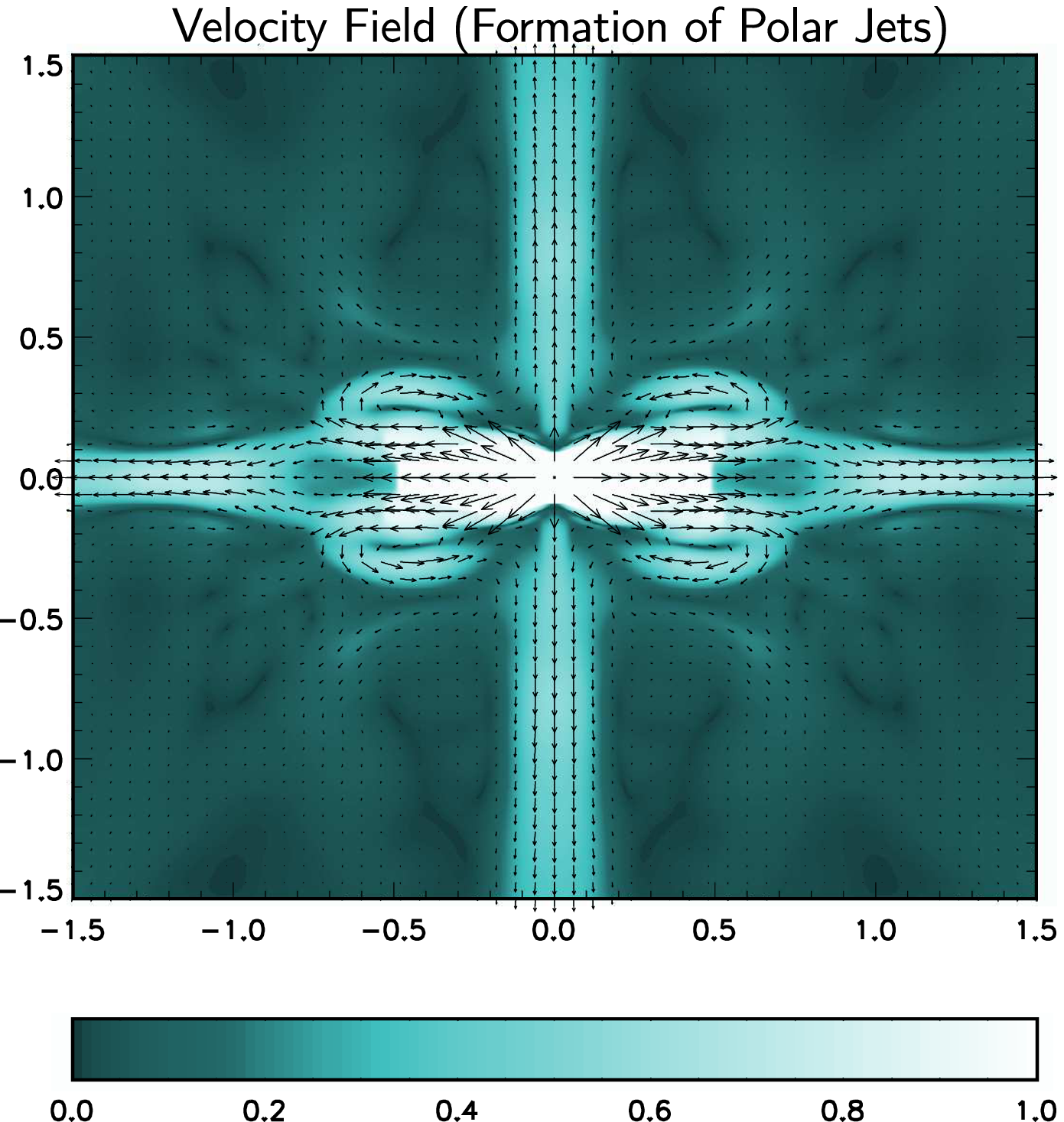} 
\caption{Maps of flow velocity (in units of $c$) from the axisymmetric simulations in \cite{Del-Zanna:2004}. In the left panel the flow structure around the TS, here for an early time $t=100$~y, is shown for a low-magnetization run without the striped-wind equatorial region ($\sigma_0=0.003$, $b\to \infty$; labels are described in the text). In the right panel we show the flow structure arising at later times for higher magnetizations and for a case with a (narrow) striped wind region (Run A: $\sigma_0=0.03$, $b=10$, here $t=300$~y). Notice the simultaneous presence of supersonic equatorial flows and polar jets.}
\label{fig:flow} 
\end{figure}

When even a narrow striped-wind region is allowed and the magnetization is high enough, a strong equatorial flow forms and part of it is readily diverted towards the polar axis by \emph{hoop stresses} due the enhanced toroidal magnetic field in the post-shock nebula, as predicted analytically \cite{Bogovalov:2002,Lyubarsky:2002} and observed in the first relativistic MHD simulations \cite{Komissarov:2003,Komissarov:2004a}. The new flow pattern is shown in the right panel, for the so-called Run A parameters ($\sigma_0=0.03$, $b=10$). We clearly see the simultaneous presence of supersonic outflows both in the equatorial plane and along the polar axis, in the form of narrow jets (or plumes). The complex pattern around the TS is expected to produce Doppler-boosted features when non-thermal emission is computed on top of the MHD quantities, and the presence of the fast outflow will advect the emitting particles at large distances both in the equatorial plane and along the polar axis.

In order to verify whether the flow pattern arising from simulations with a moderate wind magnetization $\sigma_0$ is really the correct counterpart of the jet-torus structure observed in the X-rays \cite{Weisskopf:2000}, a detailed calculation of the non-thermal emission based on the evolved MHD quantities and particle tracers must be then performed. This was achieved in \cite{Del-Zanna:2006}, where various aspects of the non-thermal emission, computed following the recipes summarized in the previous section, were discussed. Here we report, in Fig.~\ref{fig:emission}, the optical and X-ray simulated surface brightness maps for a set of parameters that match the Crab nebula emission properties at best within 2-D modeling (Run A). A square box of 6~ly $\times$ 6~ly is displayed, with intensity in logarithmic scale, normalized against the maximum in each band.

The optical image (left panel) appears to be elongated and rather uniform around the polar axis up to distances of $2-3$~ly. No hints of an equatorial torus of enhanced emission are found, while Doppler-boosted arcs, the variable wisps, and a very bright knot are present close to the pulsar position, as observed in HST data \cite{Hester:2002}. These are due to a higher value of the magnetic field close to the TS and to the presence of mildly relativistic velocities pointing towards the observer direction, as discussed in \cite{Komissarov:2004a}. The knot appears to be located at the base of the polar jet, a candidate site for the $\gamma$-ray flares observed in the Crab nebula \cite{Komissarov:2011,Yuan:2015,Lyutikov:2016}. 

\begin{figure}[t]
\centering
\includegraphics[height=55mm]{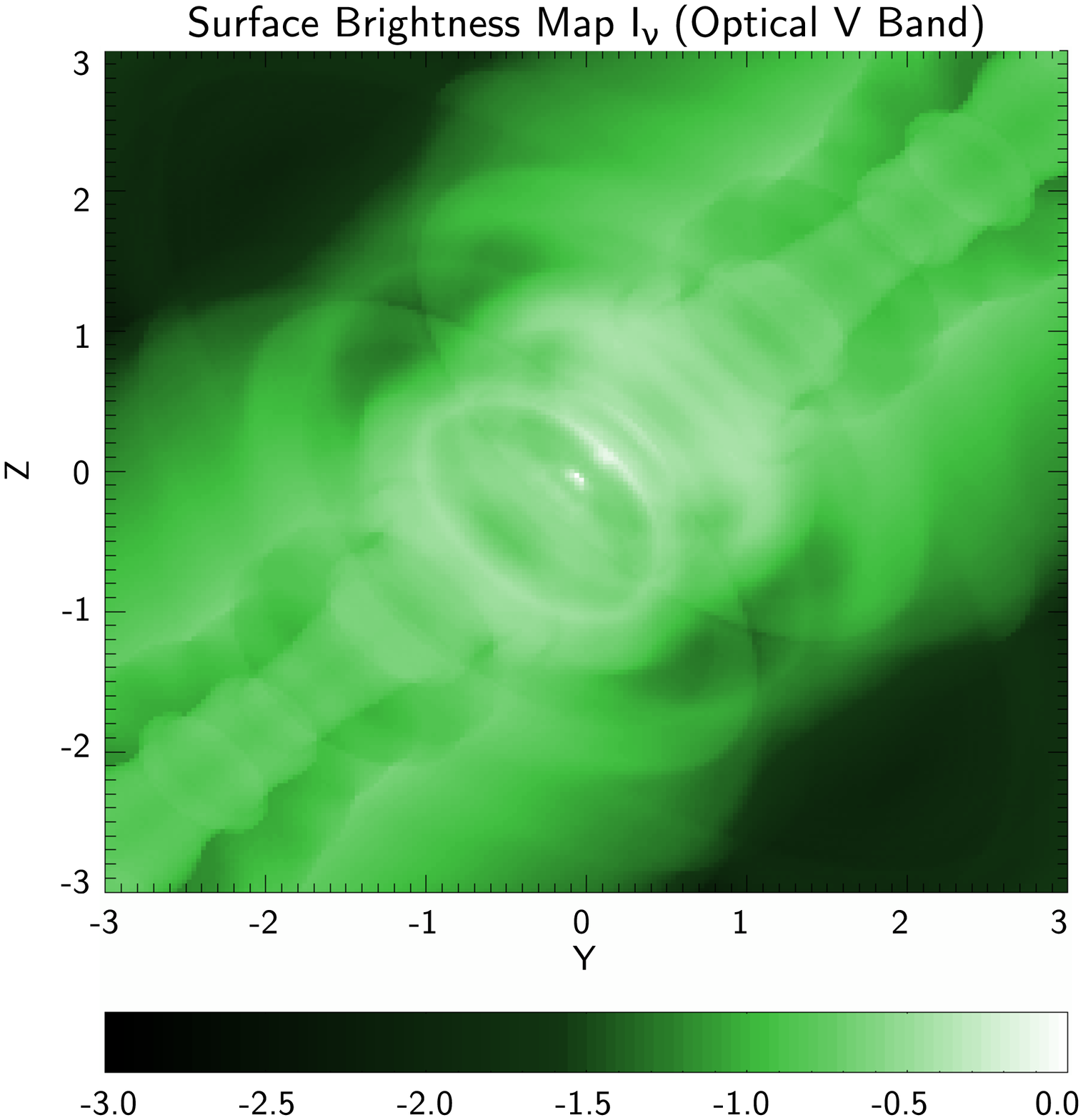} \quad
\includegraphics[height=55mm]{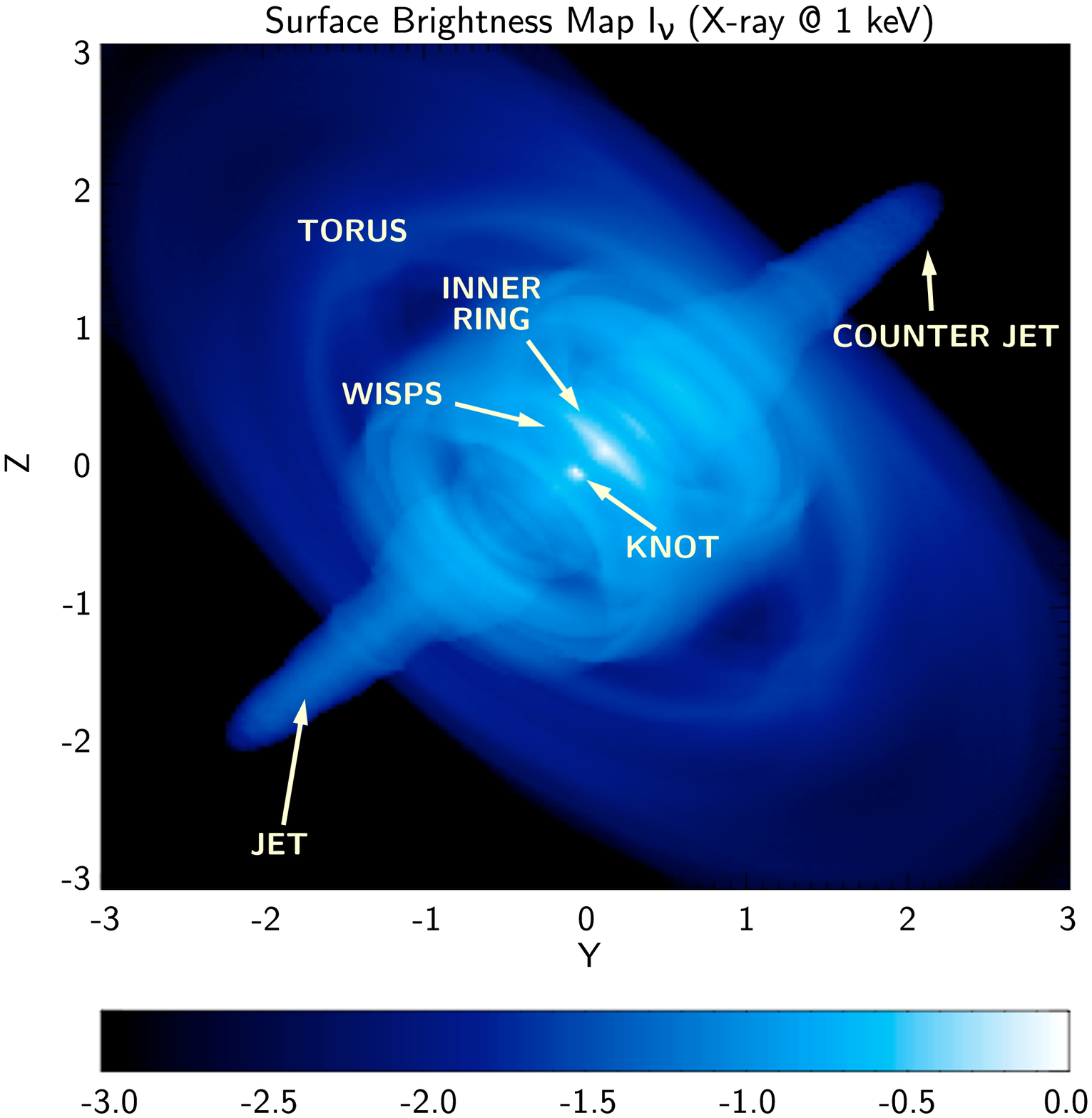} 
\caption{Simulated non-thermal emission from an axisymmetric relativistic MHD model of the Crab nebula. In the left (right) panel the optical (X-ray at 1~keV) surface brightness map is shown, taken from \cite{Del-Zanna:2006}. A logarithmic scale is employed (the intensity is normalized to its maximum) and distances from the central pulsar on the axes are measured in ly. In the X-ray image many features which are also present in real data can be identified.}
\label{fig:emission} 
\end{figure}

Due to the shorter synchrotron lifetime of higher-energy particles, the inner jet-torus structure is instead clearly visible in the X-ray surface brightness map, which is indeed very similar to what observed, at least at a qualitative level. A polar (hollow) jet and an equatorial (rather faint) torus are found, due to the injection of fresh particles advected by the flow. Both features are less bright than expected for the Crab nebula, due to a lower magnetic field (recall that the assumption of a purely toroidal field in the axisymmetric case requires $\vec{B}=0$ along the polar axis). However, the field is strong near the TS, and several arcs of brighter emission are seen, even more clearly than in the optical band. 

Time variability of the high-energy emission and of wisps in particular has been also investigated \cite{Volpi:2008c,Camus:2009,Olmi:2015}. This motion is induced by the turbulent flow structure around the TS, where vortexes are continuously created and advected away. The equatorial outflow with $v\sim 0.5 c$ is highly inhomogeneous and regions with different plasma magnetization are crossed, leading to the variable synchrotron emission observed in the form of expanding wisps. The period corresponding to the largest scale of the turbulent flow is $1-2$~y, matching what is observed for the moving features in the Crab nebula \cite{Hester:2002}. In the next section we will see how the motion of wisps can be used as a probe for the mechanisms and sites for particle acceleration in PWNe. An example of wisps as observed in synthetic X-ray maps and of the turbulent plasma, responsible for their appearance as quasi-periodical bright features inside regions of enhanced magnetic field, will be shown in Fig.~\ref{fig:wisp_map}.

The main parameters characterizing the PW, such as the degree of anisotropy, the magnetization, and the inclination of the pulsar magnetosphere (or equivalently the width of the striped wind region), are of course expected to be different for different sources, so that a detailed modeling based on 2-D MHD simulations can be carried out by varying the mentioned parameters. For instance, a wider striped wind region seems to be more appropriate for the Vela PWN, since the jet-torus structure is less apparent and two boosted arcs appear in the synthetic surface brightness maps \cite{Del-Zanna:2006}. By assuming wind prescriptions based on 3-D force-free models of the pulsar magnetosphere \cite{Tchekhovskoy:2016}, the morphology of the Vela nebula was also reproduced in a recent work \cite{Buhler:2016}, still based on axisymmetric simulations, though limited to a much shorter evolution time. This was achieved by choosing a very strong magnetization $\sigma_0 = 3$ and an inclination angle of $45^\circ$. In the same work, the modeling of the Crab nebula requires a much lower magnetization, regardless of the inclination angle assumed, namely the same $\sigma_0 = 0.03$ adopted in \cite{Del-Zanna:2006}.

\section{Probing particle acceleration mechanisms}

PWNe are known to be among the most efficient cosmic accelerators, given that the measured synchrotron $\gamma$-ray spectrum of the Crab nebula and similar sources can only be explained by emitting particles (pairs) which are accelerated up to PeV energies \cite{Arons:2012}. One-zone or radial 1-D models suggest that at least two families are responsible for the non-thermal emission: the so-called \emph{radio particles} of lower energies, with a distribution $f_0(\varepsilon_0)\propto \varepsilon_0^{-p}$ of rather flat a power-law index $p\sim 1.5$, and a family of higher energy leptons with $p\sim 2.2$, the \emph{X-ray particles} \cite{Kennel:1984a,Atoyan:1996}. The two families appear to be smoothly connected around a particle energy of $\sim 100$~GeV, which roughly corresponds to photon frequencies in the optical band for the inferred values of the nebular magnetic field of $200\mu G$ \cite{de-jager:1992}, and in some models even a single family of accelerated particles in the form of a broken power-law is able to reproduce the entire spectrum of different sources, see e.g. \cite{Bucciantini:2011}. 

A self-consistent and combined treatment of the synchrotron and IC radiation components based on relativistic MHD simulations is a very powerful diagnostic technique for PWN modeling, since strong constraints can be placed on the model parameters, like the wind anisotropy, its magnetization, or even the composition (possible presence of ions) \cite{Amato:2014}. Unfortunately, this task has been attempted so far on top of multidimensional simulations only in \cite{Volpi:2008c}, where the recipes for non-thermal emission were extended to the IC mechanism, including the effects of a Klein-Nishina differential cross-section, and different targets were considered (synchrotron, infrared, and cosmic microwave background photons). 

The multi-band spectrum (the spectral energy distribution $\nu L_\nu$) obtained in the cited paper is reported in Fig.~\ref{fig:spectrum} (right panel) for the Run A parameters, described in the previous section as those producing the surface brightness maps more similar to the images of the Crab nebula in the X-ray. The separate contributions to the synchrotron emission produced by the two families (power laws with exponential cutoff) of electrons are plotted, together with a thermal component needed to model infrared data (emission by local dust). In order to match the synchrotron X-ray and $\gamma$-ray data, a steep index of $p\sim 2.7$ (compared to stationary 1-D models) was found to be required for the higher energy particles accelerated at the TS. On the other hand, the electrons responsible for the radio emission could be taken as homogeneous in the whole nebula and constant in time. With these prescriptions, the spectral break observed in the Crab nebula's spectrum between the infrared and the optical bands naturally arises as due to the different spectral indexes assumed for the two distribution functions, separated between 100 and 300~GeV. A second break, at ultraviolet frequencies, and the further steepening of the spectrum, are instead due to burn-off effects.

\begin{figure}[t]
\centering
\includegraphics[scale=0.42]{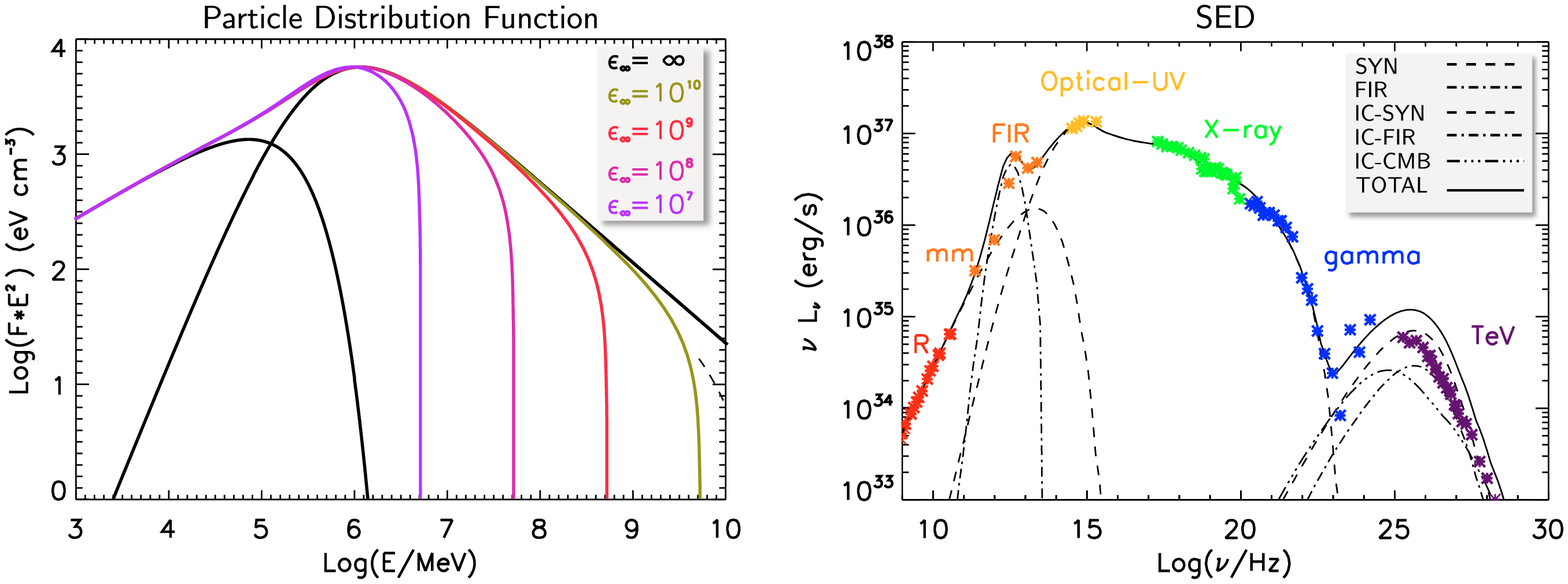} 
\caption{Simulated non-thermal emission from an axisymmetric relativistic MHD model of the Crab nebula, taken from \cite{Volpi:2008c}. In the left panel the two \emph{initial} ($\epsilon_\infty\to\infty$) energy distribution functions are shown as functions of $E=\epsilon \, m_\mathrm{e}c^2$ (dashed black lines), and the total evolved distribution function is shown for various values of the burn-off energy $\epsilon_\infty$ (see Eq.~\ref{eq:infty}). In the right panel, the resulting photon spectral energy distribution (SED) function is plotted and compared with observational data. The separated contribution to the synchrotron emission by the two families is again indicated by dashed lines, whereas the dot-dashed line refers to thermal dust emission. The contributions from different incident photon targets to the IC emission are reported with different colors and line-styles.}
\label{fig:spectrum} 
\end{figure}

The steep index required for the optical/X-ray particles is needed to compensate the reduced losses and consequent spectral softening produced by a lower magnetic field, compared to that predicted by one-zone or 1-D models. It was pointed out that, in axisymmetric simulations, the toroidal field is in fact compressed artificially around the TS and towards the polar regions, so that a small value for the average nebular field is found ($\sim 50-100\mu G$), while previous models aimed at reproducing the full spectrum indicate values which are 2-3 higher. At the same time, the number of emitting particles continuously injected at the TS must be enhanced up to values corresponding to an unnatural 100\% of acceleration efficiency from the PW, and this reflects the fact that the computed IC emission is 2-3 times higher than observed. The $\gamma$-ray emission for TeV photon energies is found to be dominated by the self-synchrotron component, while minor contributions arise from the cosmic microwave background and the thermal dust emission, in agreement with radial or one-zone models \cite{Atoyan:1996,Meyer:2010}. 

In spite of the wealth of data concerning the Crab nebula and other PWNe, the origin of the two families is still debated. Radio particles are dominant by number, and if these are continuously originated inside the pulsar's magnetosphere via cascading processes and injected in the relativistic wind first and then in the nebula, an accurate modeling of PWNe emission is expected to provide precious information on their rate of production. As mentioned in the introduction, this has important consequences also on fundamental physical issues, as pulsars are considered to be the most efficient antimatter factories in our Galaxy, and thus they may provide an astrophysical explanation to the measured cosmic ray positron excess \cite{PAMELA-coll.:2009,AMS-02-coll.:2013,Serpico:2012}. 

Going into deeper details, it is well known that particles constituting the wind and the PWN are not those directly extracted from the neutron star surface, but rather they are originated in a cascade process inside the magnetosphere greatly increasing their number by an unknown \emph{multiplicity} parameter $\kappa$. Estimates of this number crucially depend on whether the radio particles are part of the pulsar outflow or not. In the first case, stationary 1-D MHD models predict $\kappa\sim 10^4$ \cite{Kennel:1984,Atoyan:1996}, while if radio particles have a different origin, time-dependent one-zone models predict $\kappa\sim 10^6$ \cite{Bucciantini:2011}. In this latter case, the possible presence of ions in the wind would be irrelevant from an energetic point of view. It is thus clear that discriminating between the two scenarios is of fundamental importance for a correct modeling of the PW \cite{Amato:2014}.

In a recent study \cite{Olmi:2014}, different possibilities for the origin of radio particles have been considered, testing whether it is possible or not to discriminate their origin based on the resulting emission morphology. The question is thus whether radio particles can be considered as part of the pulsar outflow, and accelerated at the TS together with high energy emitting particles, or if they have a different origin and are accelerated somewhere else in the nebula. In the latter case, corresponding to either a primordial burst (relic population, still present thanks to the very long lifetime against synchrotron losses) or to a possible continuous re-acceleration, for example in the external thermal filaments, radio particles can be considered as a steady-state and uniform population. Relativistic MHD axisymmetric simulations were used to test the various scenarios, and results are discussed in detail in the cited paper. Surface brightness maps in the radio band, calculated for the first time based on numerical simulations, appeared to be undistinguishable in the two cases, thus unfortunately no constraints could be placed on the origin of radio particles and on the value of the pair multiplicity parameter. The only scenario that could be rejected was that of a primordial burst of radio particles followed by pure advection, while a continuous replenishment of the nebula seems to be required.

Another important result found in \cite{Olmi:2014} is that the appearance in simulated radio maps of bright and time-dependent features around the TS is basically independent on the adopted scenario. These features are named wisps when observed in the optical and X-ray bands, as previously discussed, and they were also found to be present in radio in the Crab nebula \cite{Bietenholz:2004}. Long-term variability of radio wisps (a few years) was measured up to $\sim 3$~ly from the central pulsar, where the inferred velocity is $\sim 0.002\, c$, whereas variability on a month scale and much quicker wisps with $\sim 0.4\, c$ were found in the equatorial torus. Simulations fully reproduce these observations, without invoking different origins for the two classes of radio wisps, thus confirming, on the one hand and once more, the validity of the fluid/MHD description, and on the other hand that these variable features simply trace the underlying plasma flow. 

\begin{figure}[t]
\centering
 	\includegraphics[height=50mm]{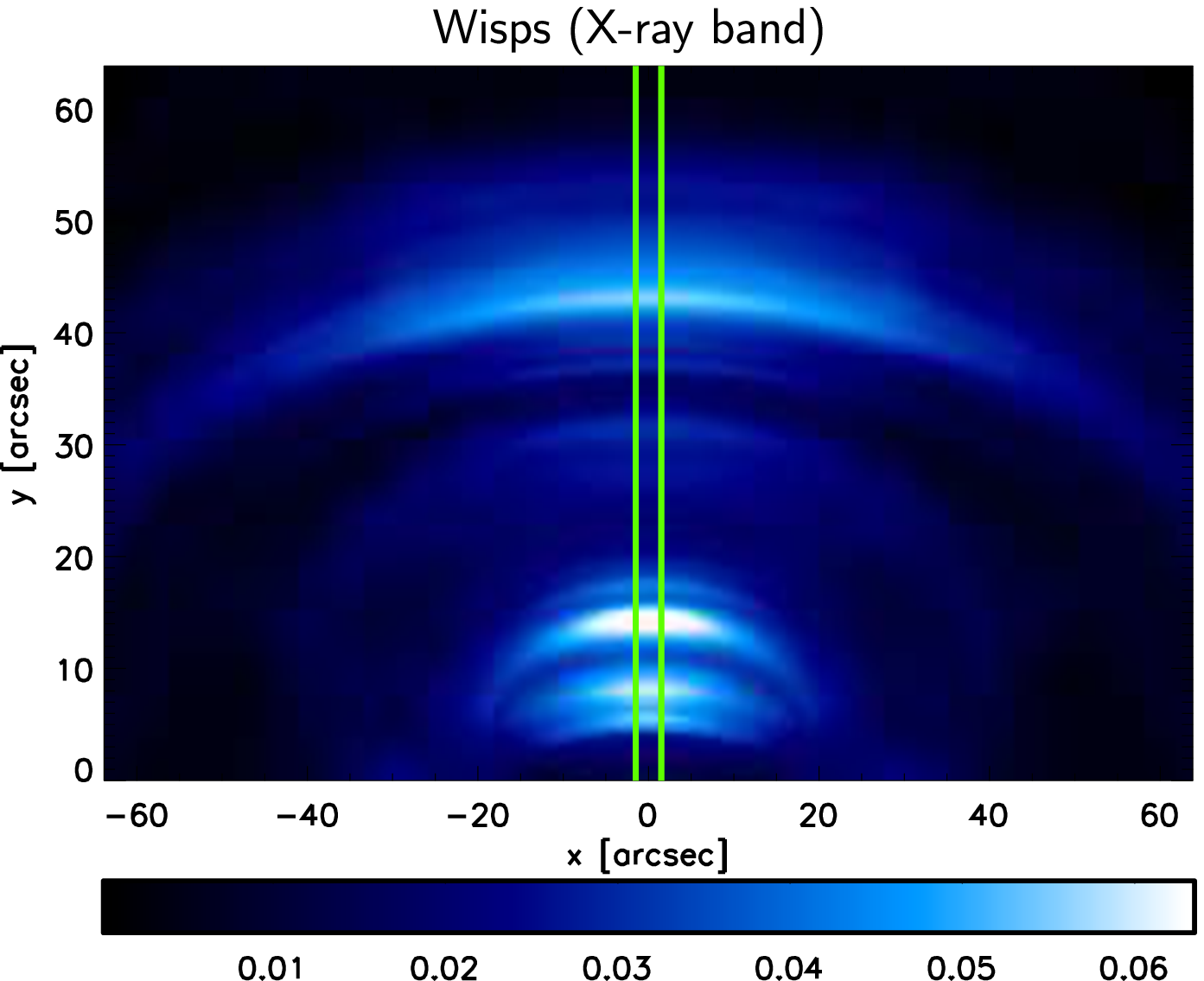} 
          \includegraphics[height=54mm]{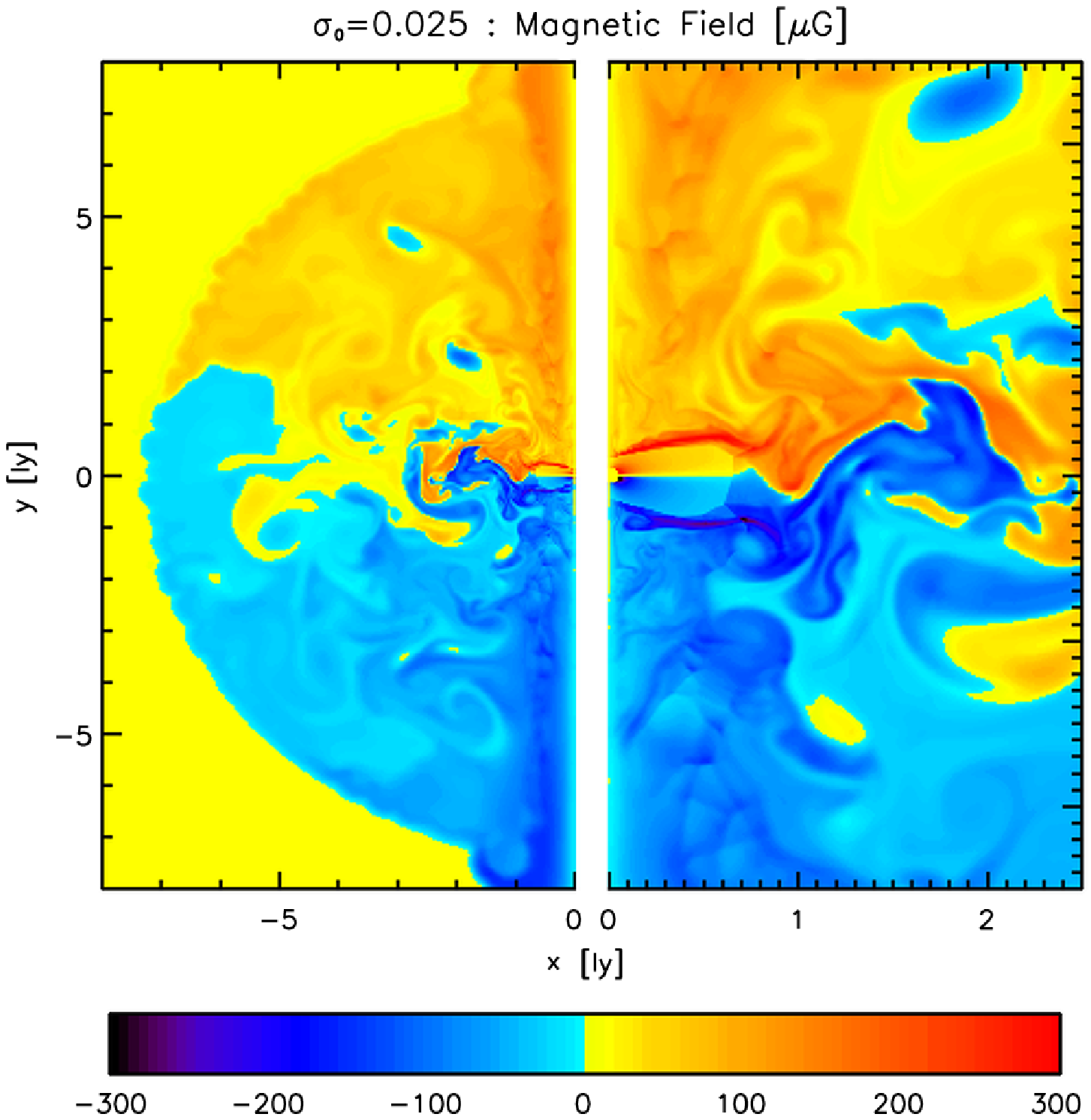}
\caption{Left panel: the simulated synchrotron X-ray surface brightness map at 1~keV, convolved with the \emph{Chandra} point-spread function, in linear scale and expressed in units of mJy~arcsec$^{-1}$, taken from \cite{Olmi:2015} (only the upper hemisphere is shown, the vertical lines indicate the sector used for extraction of the radial profiles). Right panel: the turbulent magnetic field in the whole nebula and around the TS, responsible for the quasi-periodical wisp production, taken from \cite{Olmi:2014}.
} 
\label{fig:wisp_map}
\end{figure}

However, wisps in the inner region of the Crab nebula are actually observed in (varying) positions which are not precisely coincident at the different wavelengths of observation, from radio to X-rays \cite{Bietenholz:2004,Hester:2002,Schweizer:2013}, suggesting in turn a difference in the acceleration sites of the particles responsible of such emission. Different acceleration mechanisms require very different physical conditions to be effective, and thus identifying the location at which particles have been accelerated can put strong constraints on the mechanism at work. The motion of wisps is then expected to be a powerful diagnostic tool to investigate whether radio and optical/X-ray particles are accelerated in the same site or not. 

In another recent numerical work by the same authors \cite{Olmi:2015}, different possibilities were considered for injection of particles: in case \emph{I} all particles are injected uniformly at the shock surface; in case \emph{II} particles of one family are injected in a wide equatorial sector or in the complementary narrow polar one; in case \emph{III} particles of one family are injected in a narrow equatorial region, or in the complementary wide polar one. The entire evolution of the Crab nebula was then simulated by introducing as many numerical tracers as necessary in order to follow the evolution of particles injected at the different locations. Following the analysis by \cite{Schweizer:2013}, intensity peaks were extracted from a $3^{\prime\prime}$ wide slice around the polar axis in the upper hemisphere of each surface brightness map, where wisps are more prominent (see Fig.~\ref{fig:wisp_map}}).

\begin{figure}[t]
\centering
 	\includegraphics[scale=0.65]{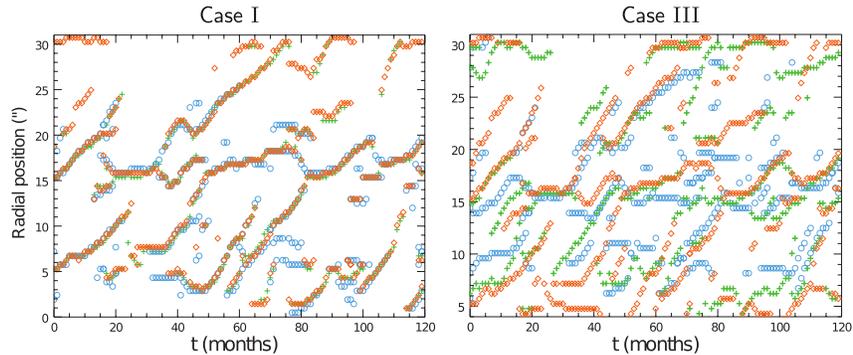}
\caption{Radial positions of the local intensity maxima (in arcsec) as a function of time (in months) with orange diamonds identifying radio wisps (at 5~GHz), green crosses optical ones (at $3.75 \times 10^{14}$~Hz) and light-blue circles for X-rays (at 1~keV).  On the left case \emph{I} is shown. On the right case \emph{III} is shown, with X-ray particles injected in a narrow equatorial zone and radio ones injected in the complementary sector. Further details in the text and in \cite{Olmi:2015}.} 
\label{fig:wisp_cases}
\end{figure}

In Fig.~\ref{fig:wisp_cases} wisps obtained in the case of uniform injection (case \emph{I}) or injection of X-ray particles in a narrow equatorial region and radio particles in the complementary sector (case \emph{III}) are compared. As expected, when particles of different families are injected at the same location, i.e. across the entire shock surface, wisps appear to be coincident at the different wavelengths. But as soon as they are injected at distinct locations, wisps at radio, optical and X-ray frequencies are no more coincident. The main result of the analysis is that, in order to reproduce the absence of X-ray wisps in the region within $\sim 6^{\prime\prime}$ from the pulsar, as observed in \cite{Schweizer:2013}, X-ray particles must be injected in a narrow equatorial sector (the case displayed in the right panel), approximately coincident with the striped region of the wind where dissipation of magnetic field is expected to be most efficient. This may indicate that X-ray particles are produced via Fermi I acceleration in the striped zone, where magnetization can be lowered enough by dissipation so as to allow this mechanism to be effective \cite{Sironi:2009}. The same conclusion was reached by the analysis of wisps motion in 3-D simulations \cite{Porth:2014}.

Moving to lower frequencies, strong constraints on radio emission are again difficult to derive: the only case that can be easily excluded is the one in which radio particles are injected in a narrow polar cone, since no wisps appear to be produced, while even a steady and uniform distribution of radio emitting particles throughout the nebula does not change the overall picture. A possible interpretation of this finding is that radio particles could result from driven magnetic reconnection primarily occurring at moderately high latitudes, where conditions for driven magnetic reconnection to operate as an acceleration mechanism might be locally satisfied \cite{Sironi:2011}, or even that radio particles are uniformly distributed in the nebula, possibly due to some continuous and ubiquitous re-acceleration mechanism, for instance induced at the Rayleigh-Taylor fingers protruding throughout the nebula \cite{Komissarov:2013}.

\section{From axisymmetry to full 3-D: solution to the $\sigma$-problem?}

The so-called $\sigma$-problem, or $\sigma$-paradox, has puzzled theorists since the very beginning of MHD modeling of PWNe. In addition to the latitudinal dependence of the wind luminosity, needed to provide the jet-torus structure of the PWN arising from the confinement of the wind itself, also the magnetization $\sigma$ of the PW is expected to vary (by several orders of magnitude) with the radial distance: at the pulsar magnetosphere the relativistic wind is certainly Poynting-dominated ($\sigma\gg 1$ in Eq.~\ref{eq:sigma}), whereas at the TS, something like a billion of light cylinder radii farther out, steady-state radially symmetric models predict $\sigma_0 = \mathrm{few} \times 10^{-3}$ to match the PWN expansion (see Eq.~\ref{eq:ratio} and comments below). An initial conversion of energy from the electromagnetic field to the bulk inertial flow is expected to occur during the acceleration process within the fast-magnetosonic point \cite{Kirk:2009},  whereas beyond that radius, up to the TS,  both $\sigma \equiv \sigma_0\gg 1$ and $\gamma\gg1$ should remain basically constant, at least for relativistic MHD radial models.

For non-aligned rotators, that is when the magnetosphere of the pulsar is inclined with respect to the rotation axis, the equatorial current sheet where the toroidal radiation-like field reverses its sign wobbles and produces a striped-wind region of reconnected field and lower magnetization, as already discussed. Its (half) angular extent is given by the pulsar inclination angle, and the periodicity of this large-amplitude wave is that of the pulsar rotation \cite{Coroniti:1990}. Detailed models of energy conversion within the striped wind region have been proposed, though it is not clear wether the rate of injection of pairs in the wind and the efficiency of the dissipation mechanism (e.g. the tearing instability) are enough to reach a condition of $\sigma_0\ll 1$ at the TS \cite{Lyubarsky:2001,Kirk:2003}. Additional magnetic dissipation can also arise at the shock location, due to kinetic effects arising when the Larmor radii of particles become comparable with the TS size itself \cite{Lyubarsky:2003,Sironi:2011}. In any case, even if magnetic to kinetic energy conversion can be efficient around the equator, certainly the polar regions are expected to remain Poynting-dominated, thus leaving the $\sigma$-paradox mostly unsolved \cite{Komissarov:2013}.

A hint towards the solution comes from the relativistic MHD simulations object of the present review. We have already discussed that when 1-D radial models are abandoned in favor of 2-D axisymmetric ones, the lower tension of the post-shock toroidal field allows to reproduce the observed jet-torus inner structure of PWNe with a moderate magnetization $\sigma_0\sim \mathrm{few} \times10^{-2}$. However, the magnetic field assumed in these models is purely toroidal, and subject to (slow) diffusion only at the equatorial current sheet. Moreover, in axisymmetric models the nebular magnetization is expected to remain low at all radii, otherwise the pinching effect would create excessive axial elongation \cite{Begelman:1992}. Indeed, 2-D simulations show a pile up of magnetic field around the TS and along the poles. But how realistic is the assumption of a purely toroidal field? 

The stability of such MHD structure (the \emph{Z-pinch} equilibrium in controlled fusion literature) was investigated by \cite{Begelman:1998} in the context of PWNe, who found that it would be very likely destabilized by 3-D kink-type instabilities very close to the TS radius. In particular, magnetized jets endowed with toroial fields are subject to these current-driven instabilities, as confirmed by relativistic MHD simulations \cite{Mizuno:2011,Mignone:2013}, and as even seen in X-ray images of the Crab and Vela PWNe, where time-variable kinks in the polar jets are actually observed \cite{Weisskopf:2000,Hester:2002,Pavlov:2003}.

The first full 3-D relativistic MHD simulations of PWNe \cite{Porth:2013,Porth:2014} settled the issue. By using the same setup of axisymmetric runs, the authors were able to produce models with $\sigma_0 = 1 - 3$, that is up to a thousand times the magnetization employed in radial MHD models. The inner jet-torus structure is preserved, but jets are now subject to kink instabilities and a high level of magnetic randomization and dissipation occurs in the main body the PWN, simply due to the extra channels of dynamics made possible by the 3-D setup. The size of the TS is correctly retrieved and it is basically the same as in lower magnetization models, thus confirming Begelman's idea that lower dimensionality models are not adequate to put constraints on $\sigma_0$, due the excessive rigidity of purely toroidal fields. PWNe can thus be modeled by assuming $\sigma_0>1$ at the TS, and the long-standing paradox appears to be finally solved.

Is this the end of the story? Unfortunately these first 3-D simulations have been run for just $\simeq 100$~y, $1/10$ of the Crab lifetime, when the self-similar expansion phase is still far from being reached. In particular, it is not clear wether the average nebular field strength will be close to that inferred from observations at the final stage of the evolution ($\simeq 100-200 \,\mu$G), and if the synthetic non-thermal emission will match observational data in all spectral bands, including the IC component.  Moreover, magnetic dissipation is entirely of numerical origin in the ideal MHD code employed, a problem that can be partly overcome in 2-D simulations by simply increasing the resolution, but certainly not in 3-D runs using AMR, which are extremely demanding from a computational point of view.

Nevertheless, it is clear that this kind of 3-D simulations are the primary tool to reach a comprehensive description of PWNe in all their aspects. Here we report preliminary results from 3-D AMR simulations of the Crab nebula by our group \cite{Olmi:2016}, where we successfully managed to reach, if not the proper lifetime of the Crab, at least the self-similar expansion phase ($t=250$~y, corresponding to an effective age of 500~y for the given initial PWN radius). The setup of the PW was the same for 2-D models, and we chose $\sigma_0=1$ and $b=10$ (the blue curve in Fig.~\ref{fig:sigma}).

\begin{figure}[t]
\centering
 	\quad \includegraphics[scale=0.27]{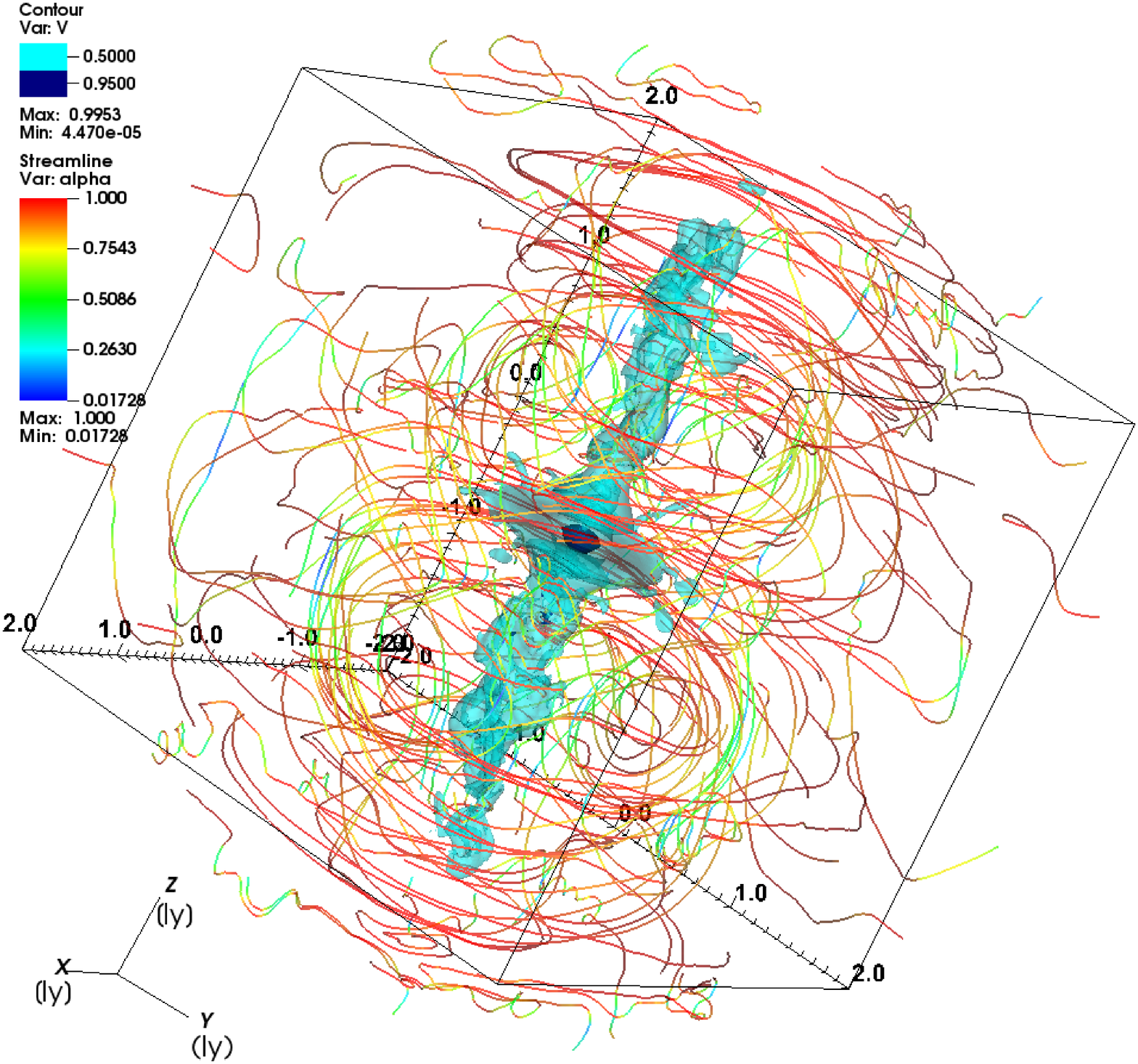}
\caption{Magnetic field lines drawn with seed points laying on a sphere of radius 2.5 ly, at the time $t=250$~y, taken from \cite{Olmi:2016}. The contrast between toroidal and poloidal components is highlighted by showing the quantity $\alpha=B_\mathrm{tor}/B$ in different colors: red means that the field is completely toroidal and blue that it is predominantly poloidal. 
The velocity field is also represented as a 3-D contour plot, with color levels corresponding to $0.95 c$ and $0.5 c$. Dimensions of the nebula are indicated by the box axes, in units of ly.} 
\label{fig:3dfield}
\end{figure}

In Fig.~\ref{fig:3dfield} the structure of magnetic field lines at $t=250$~y is shown, with the color bar indicating the ratio of the toroidal to the total field magnitude, measured by the quantity $\alpha_B=B_\mathrm{tor}/B$. As expected, the magnetic field structure is complex: the field is mainly toroidal in the inner part of the nebula, as only this component is injected in the wind and amplified post-shock, while the poloidal component is mostly important outside, in the polar and equatorial external regions, where the field structure is more efficiently modified by axial high-speed flow or by the turbulent motions, respectively. In the same figure, we also show iso-surfaces of constant velocity, dark blue indicating the wind region and light blue to highlight the jets, for which we have chosen a reference value of $v=c/2$. 

In a first phase, say up to 100~y of evolution, the average nebular magnetic field strength is found to decrease down to values of the order of $\sim 100 \, \mu$G, which means that the average field in the nebula is about a factor of 2 lower than expected. However, the trend seems to indicate that the strong magnetic dissipation of the first period of evolution slows down after that time: the field strength reaches a minimum value and afterwards stays almost constant, or even slightly increases. This might indicate that magnetic dissipation progressively becomes less important, leading the magnetic field to saturate to the equipartition value at some point of the evolution, so we are confident that the emission properties derived from this simulation can in principle match the observations.

\begin{figure}[t]
\centering
 	\includegraphics[scale=0.38]{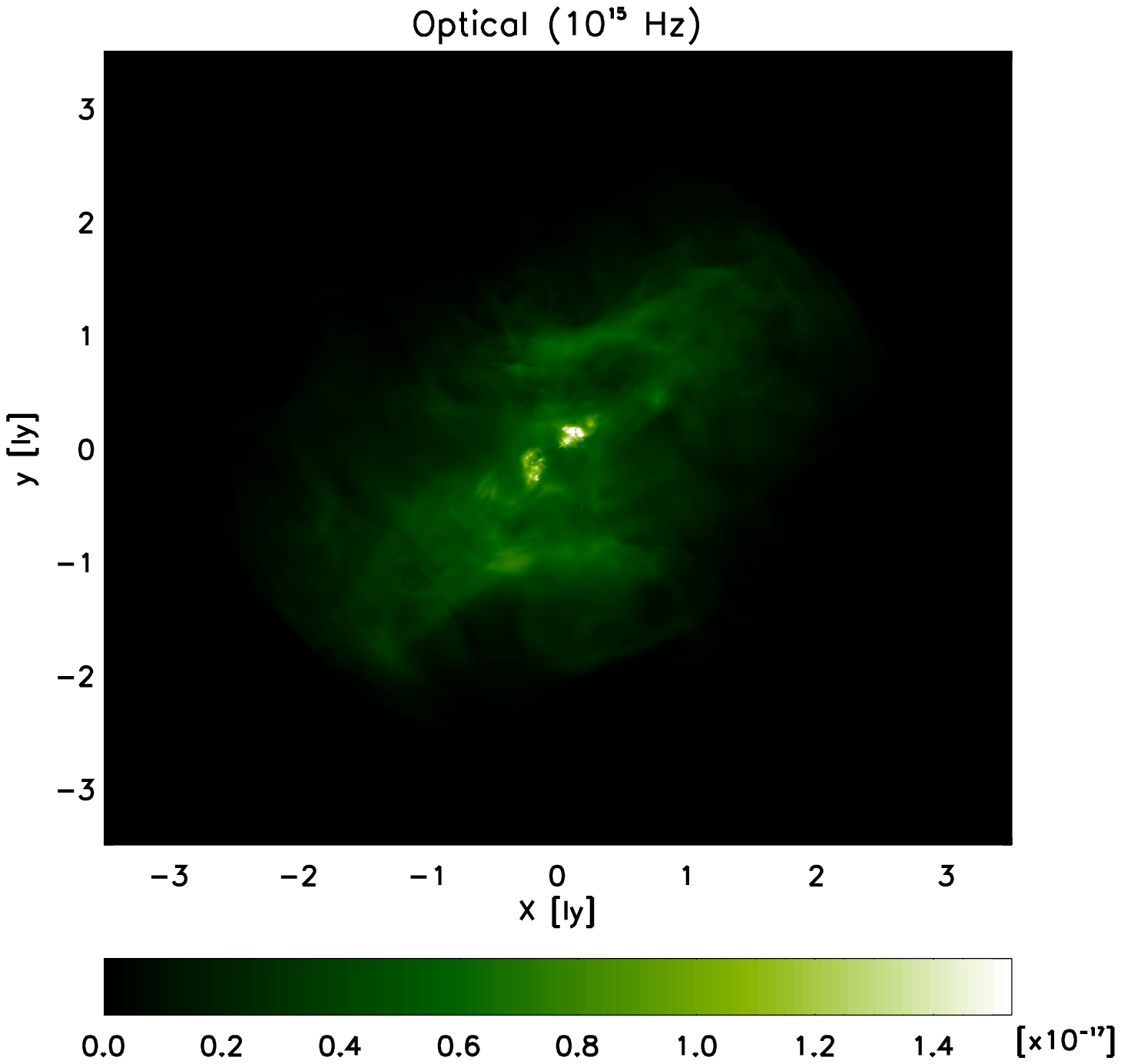}
	\includegraphics[scale=0.38]{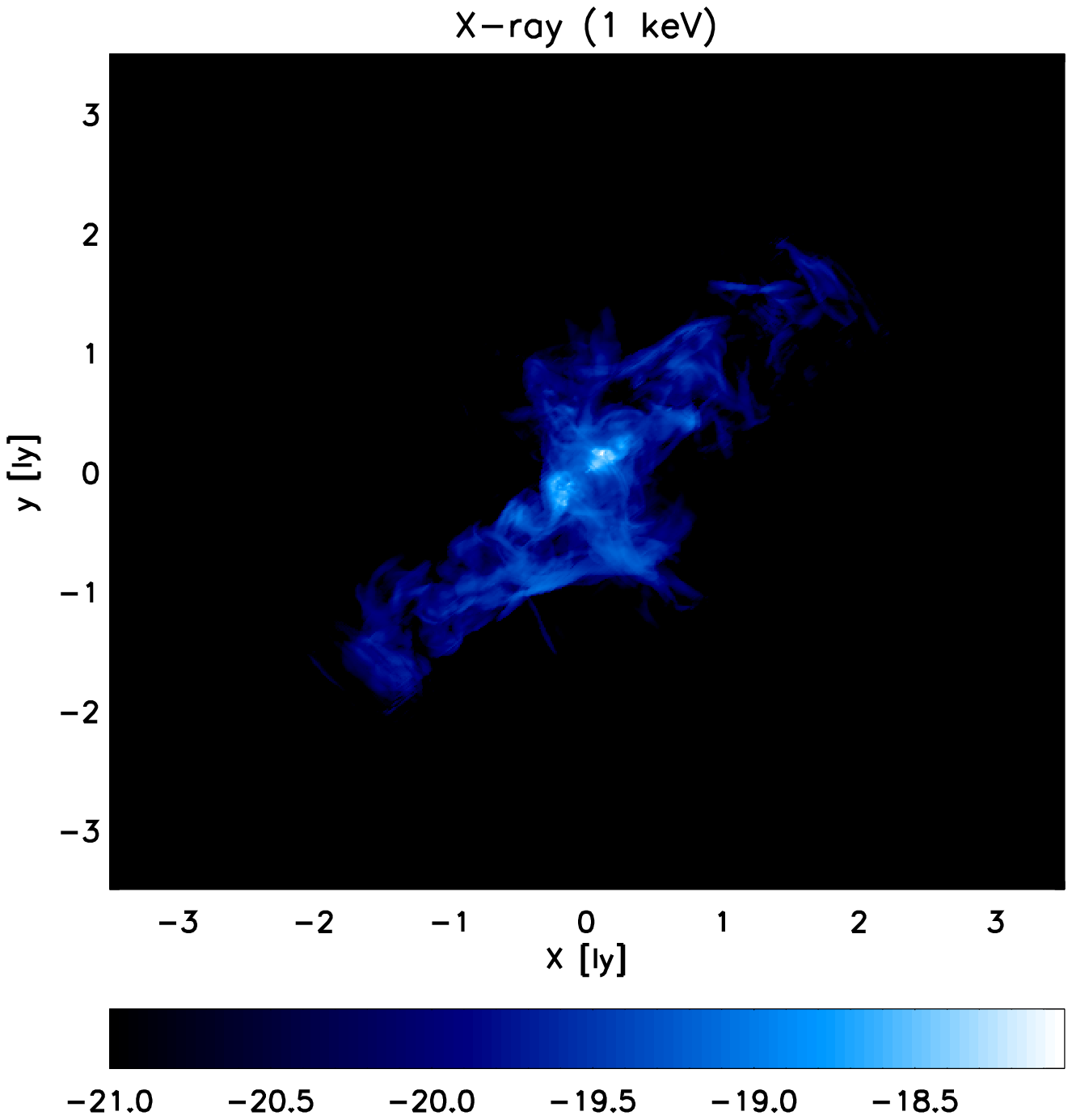}
\caption{Surface brightness maps from 3-D simulation of the Crab nebula at $t=250$~y, taken from \cite{Olmi:2016}: the optical map at  $\nu= 10^{15}$~Hz is displayed in the left panel, the X-ray one at 1~keV is displayed in the right panel. In both maps the intensity is expressed in mJy/arcsec$^2$ units and distance from the pulsar is in ly.  
	 } 
\label{fig:3dmaps}
\end{figure}

Surface brightness maps at optical frequencies and in the X-rays are shown in Fig.~\ref{fig:3dmaps}, corresponding to the end of the simulation at $t=250$~y. As can be seen, the synchrotron burn-off effect is clearly apparent at increasing photon energies, and emission maps are encouraging.  Polar jets are well formed and bright, contrary to what found in \cite{Porth:2014}, but the torus is quite under-luminous. Most of the emission comes indeed from the base of the jets, where the particle density is high, while in the torus zone, in spite of a still substantial magnetic field, the number of particles is considerably lower and the non-thermal emission is reduced.

Investigations on wisps and a study to infer their motion and outflow properties were also achieved. As expected, wisps appear to be more prominent in the upper hemisphere of the nebula (with respect to the pulsar equatorial plane) than in the lower one. Properties of wisps were obtained with the same technique as in the 2-D case, by subtracting the different intensities in emission maps taken at different times. We chose to study the variability in the radio band, for which the expected periodicity of wisps is of the order of years. A number of non-axisymmetric structures appear on the large scales, and the  difference map shows a much closer resemblance to the time variability data reported in \cite{Bietenholz:2004} with respect to the previous axisymmetric study. However, in the inner nebula, results appear to be rather similar, supporting the conclusion that the 2-D description with a dominant toroidal field is still adequate in that region.

To conclude, the first 3-D relativistic MHD simulations are promising and seem to provide a reasonable solution to the long-standing $\sigma-$paradox. The inner structure is basically unaffected by the increased dimensionality, and the condition for $\sigma_0\ll 1$ at the TS can be finally relaxed. The new channels for 3-D dynamics allow for field randomization and dissipation either in the polar jets (kink instabilities) and in the main body of the nebula. However, the final word can only come from an accurate modeling of the combined synchrotron and IC emission, with the same prescriptions for the evolution of emitting particles as (partially) successfully applied in the 2-D case. This research is currently ongoing.

\section{Summary and conclusions}
\label{sect:disc}

Multidimensional MHD simulations have been widely recognized as a powerful tool for the investigation of the physics of the relativistic magnetized plasma in PWNe. More than a decade of axisymmetric models and the recent full 3-D approach have certainly helped in the understanding of many aspects of the mechanisms at work in pulsar winds, in the associated nebulae, and those responsible for particle acceleration, though a few open questions still remain.

The turn of the millennium presented a new challenge to theorists: an unexpected \emph{jet-torus} structure was found in the X-ray emission of a number of nebulae, and this is most spectacularly observed in their class prototype, the Crab nebula \cite{Weisskopf:2000}. Axisymmetric simulations were employed to compute the flow structure and other physical quantities downstream of the TS, allowing one to reproduce the emission of PWNe at different frequencies. The synthetic surface brightness maps of synchrotron radiation obtained on top of simulations were so strikingly similar to observations, down to the very finest details, that modelers could safely conclude that relativistic MHD and axisymmetry are very reasonable approximations, and that the challenge was won!

In particular, the jet-torus structure arises as the result of the PW structure and in particular of the anisotropy of its energy flux, with a higher power injected in the equatorial region as previously predicted analytically \cite{Lyubarsky:2002}. Jets are stronger when hoop stresses are able to redirect the equatorial flow towards the polar axis, thus depending on the wind magnetization parameter $\sigma_0$ (measured at the TS).  In the innermost region, bright arcs (the \emph{wisps}) and dots (and the \emph{knot} in particular) appear as Doppler-boosted and time-variable features, determined by the highly turbulent flow structure around the TS and tracing the local magnetic field, which is amplified post-shock \cite{Komissarov:2003,Komissarov:2004a,Del-Zanna:2004,Del-Zanna:2006,Camus:2009}. 

The open problems of PWN physics, still not completely and satisfactory solved by MHD numerical modeling, are the following \cite{Olmi:2016}: the wind magnetization paradox, the origin of radio emitting particles and the related implications for the multiplicity of pairs, the precise location and mechanisms for particle acceleration at the TS, the origin of the $\gamma$-ray flares unexpectedly observed in the Crab nebula. The first three problems have been addressed here in the context of ideal MHD simulations, while the last one is likely to involve dissipative processes, namely relativistic tearing instabilities and fast reconnection, that have been studied so far just in limited numerical domains and never in global axisymmetric or 3-D simulations of PWNe \cite{Cerutti:2013,Cerutti:2014,Sironi:2014,Del-Zanna:2016}.

As far as the $\sigma$-problem is concerned, we have seen how low-dimensionality models do not allow for enough dissipation of the toroidal magnetic field inside the nebula, so that the tension remains strong and, in order to match the observed TS and PWN radii, the magnetization $\sigma_0$ at the TS can be $\sim 0.01-0.1$ at most. On the other hand, 3-D models relax this constraint: magnetic structures can kink and dissipate efficiently throughout the nebula, as predicted by \cite{Begelman:1998}, and equipartition or even conditions of  $\sigma_0>1$ can be reached in the far region of the wind \cite{Porth:2013,Porth:2014,Olmi:2016}. While the structures in the inner region are basically unchanged, since the toroidal field is still the dominant component there, the situation is drastically different in the polar jets, now prone to non-axisymmetric kink instabilities \cite{Mizuno:2011,Mignone:2013}, and far from the torus. Poloidal components are created, and new channels for efficient magnetic dissipation are opened. However, what is still puzzling is how the strongly Poynting-dominated wind originating in the pulsar magnetosphere can convert most of its electromagnetic energy into the kinetic bulk flow component, which may still be dominant in the equatorial striped wind region. Moreover, the final word on the validity of the 3-D models, to date still be run up to realistic ages, can only come from the spectral properties, from the radio up to the TeV emission by IC scattering, in order to constrain both the kinetic and magnetic content of the nebula \cite{Volpi:2008c}.

The radio emitting particles are an important component of the non-thermal energy budget, and it is fundamental to understand whether they are part of the pulsar outflow or not, but they are unfortunately challenging to be modeled. Surface brightness maps and the appearance of wisps in numerical (axisymmetric) simulations seem to be insensitive on their origin: due to the long synchrotron lifetime, radio particles can be either accelerated at the TS and advected downstream, like higher energy particles (in this case they are part of the pulsar outflow), or they can be simply modeled as a steady and uniform distribution in the whole PWN. Therefore, no conclusion on the pulsar multiplicity can be drawn at this stage. The only hypothesis that can be rejected is that of a relic population due to an early outburst, because some sort of continuous re-acceleration would be required anyway \cite{Olmi:2014}.

Finally, by investigating the motion of wisps, which appears to be slightly different at various photon frequencies, the site of particle acceleration can be in principle determined by comparing the numerical results against multi-wavelength observations of the Crab nebula \cite{Bietenholz:2004,Schweizer:2013}. Both axisymmetric and 3-D simulations seem to suggest that the best match is obtained when high-energy particles are accelerated in a narrow belt in the equatorial region of the TS \cite{Porth:2014,Olmi:2015}, whereas once again no stringent conclusion can be drawn on radio emitting particles. The equatorial region is believed to be the lowest magnetization site, if effective dissipation takes place in the striped wind, and this conclusion is also supported by the MHD modeling of the knot of the Crab nebula \cite{Lyutikov:2016}. The most natural mechanism responsible for the acceleration of those highest energy particles should be then Fermi I, which provides a particle spectral index close to that inferred from X-ray observations and it has been proven to be very effective at relativistic shocks of low enough magnetization \cite{Spitkovsky:2008,Sironi:2011}.

In spite of the success of axisymmetric simulations, the future is clearly in 3-D models. Such runs are extremely demanding in terms of computational time, since several refinement levels are required for the adaptive mesh to treat simultaneously the injection region, with the ultra-relativistic PW, as well as the external regions up to the supernova ejecta. While the first simulations reached just $\sim 1/10$ of the Crab nebula lifetime, a first attempt towards reaching the self-similar phase of expansion was recently made \cite{Olmi:2016}. A slowing down of magnetic dissipation was observed as soon as self-similar expansion begins, reaching a value for the average magnetic field of about 100 $\mu$G, only a factor 2 lower than what observed. Emission at radio, optical and X-ray frequencies was computed and the synchrotron burn-off effect was observed through the decrease of the size of the emitting area with increasing observation frequency, as in 2-D. However, the brightness contrast of the torus is rather low when compared against observations, the brightest features lying at the bases of the jets, where kinetic pressure is higher. Doppler-boosted dots and outgoing wisps are also retrieved in the innermost region, once again confirming the validity of 2-D results.

What are the next frontiers for PWN simulations? Most probably the multidimensional MHD modeling should be extended to non-ideal plasma effects, like a finite resistivity allowing for reconnection in thin current sheets \cite{Del-Zanna:2016,Landi:2015}, arising either at the TS, throughout the nebula, or inside the kinking jets \cite{Mignone:2013,Striani:2016}. Reconnection, or at least fast magnetic dissipation, is already present in simulations, though of numerical origin alone, arising whenever small scales are reached by the turbulent motions in the simulated plasma. The proper introduction of resistivity in the Ohm's law employed in relativistic MHD global simulations of PWNe would certainly represent an important upgrade.

On the other hand, one could think to improve the numerical treatment of the non-thermal particles responsible for emission, by adopting the hybrid MHD/kinetic approach first introduced for cosmic-ray acceleration at non-relativistic shocks \cite{Bai:2015} and later used to model the leptonic radiation from the magnetized plasma in AGN jets \cite{Vaidya:2016}. A first attempt has been already achieved by studying the diffusion of particles in evolved PWNe \cite{Porth:2016}, though in this work test particles are simply transported on the unperturbed MHD background provided by 3-D simulations \cite{Porth:2014}. A major breakthrough will come when the self-consistent hybrid MHD/kinetic approach for the Lagrangian test particles will be first employed to the transport and evolution of the families of accelerated leptons in PWNe, either at shock locations or in reconnecting current sheets.

To conclude, while numerical MHD modeling of PWNe has progressed impressively in more than a decade of simulations, the next years could witness newer and even more exciting achievements along the lines mentioned above. As already happened in the past, the future progresses in the numerical modeling of PWNe will surely also reflect in a better understanding of the physics of other astrophysical sources powered by magnetized relativistic outflows, most notably AGNs an GRBs.

\vspace{1cm}
\noindent
{\bf Acknowledgements}
\vspace{5mm}

\noindent
The authors sincerely thank the other members of the Arcetri high-energy astrophysics group, E. Amato, N. Bucciantini, R. Bandiera and D. Volpi, without whom most of the results reported in this review would have never been achieved. LDZ and BO also acknowledge support from INAF - Arcetri, from the INFN - TEONGRAV initiative, from MIUR (grant PRIN 2015: \emph{Multi-scale Simulations of High-Energy Astrophysical Plasmas}), and from CINECA for computational resources (\emph{CRAB3D: Three-dimensional relativistic MHD simulations of the dynamics and non-thermal emission of the Crab nebula}).

\footnotesize{
\bibliography{biblio_ldz}}

\end{document}